\documentclass[natbib,twocolumn]{svjour3}                     
\smartqed  
\usepackage{graphicx}
 \usepackage{mathptm}      
 \usepackage{times}
  \usepackage[official]{eurosym}
\def\arcsec{\hbox{$^{\prime\prime}$}}

\begin{document}

\title{Prospects of solar magnetometry  - from ground and in space
}


\author{Lucia Kleint        \and
        Achim Gandorfer  
}


\institute{L. Kleint \at
              Fachhochschule Nordwestschweiz,  Bahnhofstrasse 6, 5210 Windisch, Switzerland  \\
              \email{lucia.kleint@fhnw.ch}           
           \and
           A. Gandorfer \at
Max-Planck-Institut f\"ur Sonnensystemforschung, Justus-von-Liebig-Weg 3,
D-37077, G\"ottingen, Germany \\
             \email{gandorfer@mps.mpg.de}
}

\date{Received: date / Accepted: date}

\maketitle

\begin{abstract}

In this review we present an overview of observing
facilities for solar research, which are planned or will come to operation in near future. We
concentrate on facilities, which harbor specific potential for solar magnetometry.
We describe the challenges and science goals of future magnetic measurements, the status of magnetic
field measurements at different major solar observatories, and provide an outlook on possible upgrades of future instrumentation.


\end{abstract}

\section{Introduction: Complementary worlds - the advantages and drawbacks of ground-based and space-borne instruments}
\label{se:intro}

\begin{figure*} 
  \centering
   \includegraphics[width=.95\textwidth]{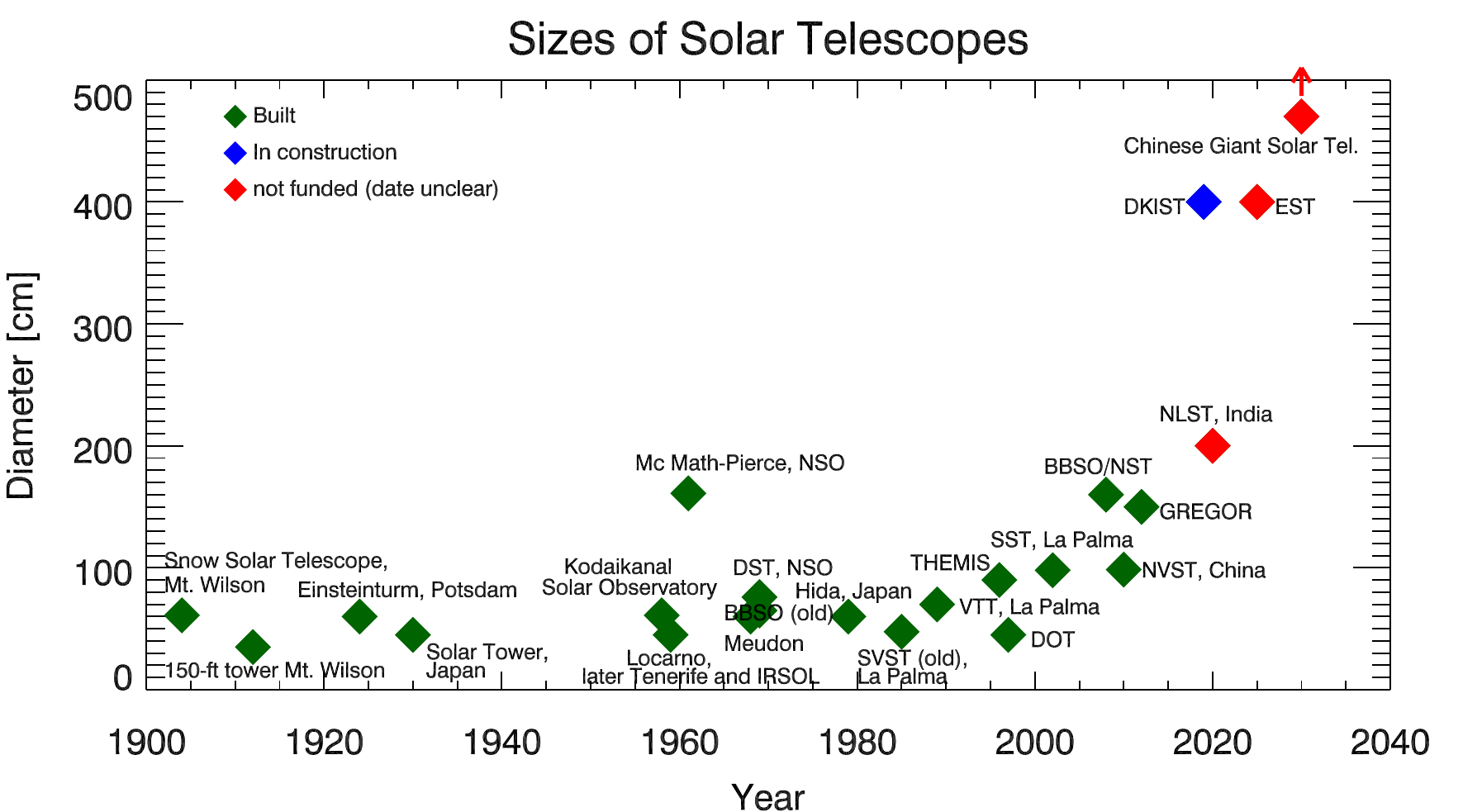}
   \caption{The effective aperture size of selected telescopes shown for the year that the telescope was inaugurated.
   The DKIST will be an impressive jump in size (4 m) compared to the currently largest BBSO/NST (1.6 m).}
        \label{telsize}
  \end{figure*}

To nighttime astronomers it usually sounds as a paradox that solar magnetic measurements are photon-starved.
Detecting four polarization states ($I$, $Q$, $U$, $V$) with high enough spatial resolution (sub-arcsecond) in a relatively large field-of-view (several dozen arcsec), in a short time (less than a second), plus in sufficient wavelengths in and around a spectral line poses strict limitations on the instrumentation. Night-time telescopes simply increase their aperture in order to collect more photons, with the currently largest aperture
of 10.4 m at the Gran Telescopio Canarias (Grantecan) on La Palma, Spain. This is also desirable for solar telescopes, but the technical requirements are more complicated owing to the heat management and the required corrections for seeing variations by adaptive optics (AO). The construction of the world's largest solar telescope, the 4 m DKIST, led by the National Solar Observatory, is currently underway with planned first-light in 2019.
Compared to the current largest solar telescope, the 1.6 m New Solar Telescope (NST) at the Big Bear Solar Observatory, the photon collecting area will increase by a factor of more than  six. A selection of effective aperture size of solar telescopes is shown in Fig.~\ref{telsize}. Only recently, with the exception of the McMath-Pierce telescope, the aperture sizes surpassed 1~m.

\subsection{The Need for Large Apertures}\label{lgap}

A large telescope serves two main purposes: to increase the number of collected photons and to increase the spatial resolution. Both obviously can be traded for another, depending on the science question that is being investigated.

There are several science questions that currently cannot be answered because of the limited available resolutions and sensitivities - temporal, spatial, and polarimetric. For example, the need for a large number of photons is evident when searching for horizontal magnetic fields, which are observed in linear polarization. These small-scale fields are important for a better understanding of the surface magnetism, e.g.~whether they are created by local dynamo processes. The answer to the long-standing question of whether small-scale magnetic fields are more horizontal or vertical seems to currently depend on the method of analysis, and especially how the noise in $Q$ and $U$ is dealt with \citep[e.g.,][]{litesetal2008,andres2009, stenflo2010, borrerokobel2011, orozcobellot2012, steinerrezaei2012, valentin2013}.
Because the signals of linear polarization ($Q$, $U$) are much smaller than the circular polarization ($V$) in the
photospheric quiet Sun, there is a bias to detect line-of-sight fields more easily. The solution would require a higher polarimetric sensitivity, which can be reached by a larger number of collected photons. In the chromosphere, the problem becomes even more complicated because the magnetic fields are generally weaker leading to even lower polarization signals, and because the selection of spectral lines is more limited,
also in terms of Land\'e factors, again leading to lower polarization signals. Simulations of the chromospheric 8542 \AA\ line have shown that a noise
level below $\sim 10^{-3.5}$ I$_c$, I$_c$ being the continuum intensity, is required to detect quiet Sun magnetic
fields 
\citep{jaimeetal2012}.
Currently, this level is rarely reached, except for observations with the Zurich Imaging Polarimeter
\citep[ZIMPOL,][]{Povel1995, Gandorferpovel1997, Gandorferetal2004, Kleintetal2011}.
ZIMPOL modulates the polarization with frequencies of several kHz,
which eliminates seeing-induced crosstalk and allows to reach polarimetric sensitivities of up to 10$^{-5}$ I$_c$.
However, this is done at the expense of spatial resolution, which generally is $>$1\arcsec\,pixel$^{-1}$, and requires averaging over a
large part of the spectrograph slit.

Another argument to aim for large numbers of photons are coronal measurements. The free energy stored in coronal magnetic fields is believed
to drive solar flares and coronal mass ejections. But so far, it cannot be measured with the desired temporal and spatial resolution.
The coronal intensity $I_{\rm cor}$ is about $10^{-5} - 10^{-6} $ I$_c$ and its polarimetric signal is even smaller: $10^{-3} - 10^{-4} $ I$_{\rm cor}$.
Currently, it takes 70 minutes of exposure time at the 0.4 m Solar-C Haleakala telescope with a huge pixel size of 20\arcsec $\times$ 20\arcsec\ to obtain
the desired sensitivity of 1 G for coronal measurements
\citep{linetal2004}.

Scattering polarization is another prime candidate for photon-starved observations. Because of its low amplitudes, usually only small
fractions of a percent of the continuum intensity, it is notoriously difficult to observe
\citep{stenflokeller1997}.
With integration times
of several minutes and averaging over most of the length of the spectrograph slit (few dozen of arcsec), one obtains a puzzling result: it
seems as if the field strength of the turbulent magnetic field depends on the spectral line that was observed. For example, measurements in the
atomic Sr~I
 line consistently give higher magnetic field values on the order of 100 G
 \citep{trujillobuenoetal2006},
 than molecular lines, such as CN ($\sim 80$ G)
 or C$_2$ ($\sim 10$ G)
 \citep{shapiroetal2011, kleintetal2011b}.
 An explanation based on modeling was proposed by
 \citet{trujillobueno2003},
 suggesting that Sr~I may be formed in intergranular lanes, where the magnetic field is stronger and C$_2$ in granules, where the field is weaker.
 But for a conclusive explanation, one would need Hanle imaging without having to average over large spatial scales. Current solar telescopes are
 unable to collect a sufficient number of photons for the required polarimetric sensitivity and spatial resolution on time scales before the granulation changes.

Flare observations would mostly benefit from an increased temporal resolution, which
again is only feasible with more photons and faster instrumentation. A scan of a spectral line with full polarimetry
takes about 15 seconds at the Dunn Solar Telescope with the IBIS instrument. The evolution and motion of flaring plasma is clearly
noticeable during this time. The Stokes vector and possible changes of the magnetic fields are then hard to interpret.

On the other hand, increased spatial resolution will also prove highly interesting to for example compare the Sun with
state-of-the-art 3D radiative MHD simulations whose resolution reaches
few km
\citep[for details see][]{rempelschlichenmaier2011},
a factor of $\sim$5 better than current observations.
Open questions about the origin of the Evershed flow, the overturning convection and small-scale features, such as
umbral-dots and penumbral filament structure may then be resolved.

In summary, one can identify many science topics, which will greatly benefit from a 4~m telescope, e.g.:
\begin{itemize}
\item Magnetic field measurements throughout the solar atmosphere (including the corona)
\item Comparisons with high-resolution simulations, including fast polarimetry at maximum resolution
\item Turbulent magnetic fields and studies of the solar dynamo, which require Hanle imaging
\item Flares, small-scale dynamics and trigger mechanisms. Space weather.
\end{itemize}

\subsection{Advantages and Disadvantages of Ground- and Space-based Telescopes}

There are many advantages of ground-based observations, one being the possibility to upgrade and to repair instruments.
The flexibility of the instrumentation and the wavelength selection, e.g. by simply replacing a prefilter, are also important.
The ``unlimited'' telemetry, which possibly is only limited by data transfer rates and the size of data storage, is another factor
when comparing e.g. the SST (several TB during a good day) with SDO, the solar space mission with the currently highest data rates,
which requires a  dedicated ground-station to download its $\sim$1.5~TB/day
\citep{sdo2012}.
Another advantage of ground-based observations
is the possibility of real-time target changes, especially important when targeting flares or rapidly changing features, for example.
Additionally, the technical possibilities to launch a 4~m telescope currently do not exist and the costs are generally much lower for
ground-based observations.



There are more or less obvious reasons for launching telescopes into space.
The most obvious one is the absorptance of wide portions of the solar spectrum by the Earth´s
atmosphere; especially in the UV, observations of the Sun (e.g. the transition region) are restricted to
space-borne instruments.
Less obvious, but of increasing interest is the capability of space
telescopes to observe the Sun from a different vantage point, offering
complementary views on our star. The Stereo mission has impressively
demonstrated the advantages of stereoscopic viewing
\citep{harrisonetal2009,aschwanden2011}.
Of particular interest are observations of the solar polar regions, which
can be obtained from the ecliptic with only marginal quality (up to 7degrees,
thanks to the inclination of the solar rotation axis). Views from inclined
orbits my greatly improve on this shortcoming. Solar Orbiter is the first
mission carrying telescopes to an orbit, which is inclined with respect to
the ecliptic by up to 34 degrees.
But even "normal" observations, which could {\it per se} be done from ground, are
worth being done from space: The fact that most solar observatories are
located in low to mid geographic latitudes excludes a 24 hour view on our
target, a severe drawback considering the dynamic timescales of solar
activity. For a limited time of up to several days, stratospheric ballooning
in high geographic latitudes essentially offers uninterrupted observations,
as demonstrated by the Flare Genesis Experiment
\citep{bernasconietal2000}
and by Sunrise
\citep{bartholetal2011,solankietal2010}.
But only space guarantees uninterrupted viewing for weeks, months, or even
years, with constantly high data quality and without the effects of seeing.

This comes, however, to a high prize, not only literally. Space instruments generally
lack the flexibility to adjust for new observing strategies (especially the choice of observed spectral bands), and are
notoriously short of telemetry, not accounting for potential loss of data due to radiation induced upsets. The need to build an instrument under severe
restrictions in terms of mass, volume, power allocation, and often in
combination with harsh radiation and varying thermal environment makes it
necessary to agree on technical compromises, which often limit the performance of
the instrument. Only thanks to the unrivaled stability of the observing
conditions, these instruments can - for specific observations - outperform their ground-based
counterparts.
A drawback for space-based missions is their long lead time and
often their electronics, such as cameras, are no longer state-of-the-art by the time they
launch.
Because space instruments generally have a fixed configuration and often are simpler in design than their
ground-based counterparts in order to be fail-safe, data pipelines for space-based instruments are better developed and
more stable. The importance of this issue has been recognized and there currently is an effort in the DKIST project to
develop such pipelines for ground-based projects as well.
In general, ground- and space-based observations ideally complement each
other, and this will always be the case.
The question is not, which  route - ground or space - to take. The key point
is how to make optimum use of the complementary aspects of both worlds.

\section{Current and future projects in ground-based solar physics}
In this section, we review a selection of current and future optical telescopes from all over the world and their measurements of magnetic fields. Access to observing time of most European telescopes (and to the DST) is possible through the SOLARNET consortium, a European FP7 Capacities Programme. Access to US-based publicly funded telescopes (DST and later DKIST) is possible through an open proposal process.

\begin{figure*} 
   \centering
    \includegraphics[width=.8\textwidth]{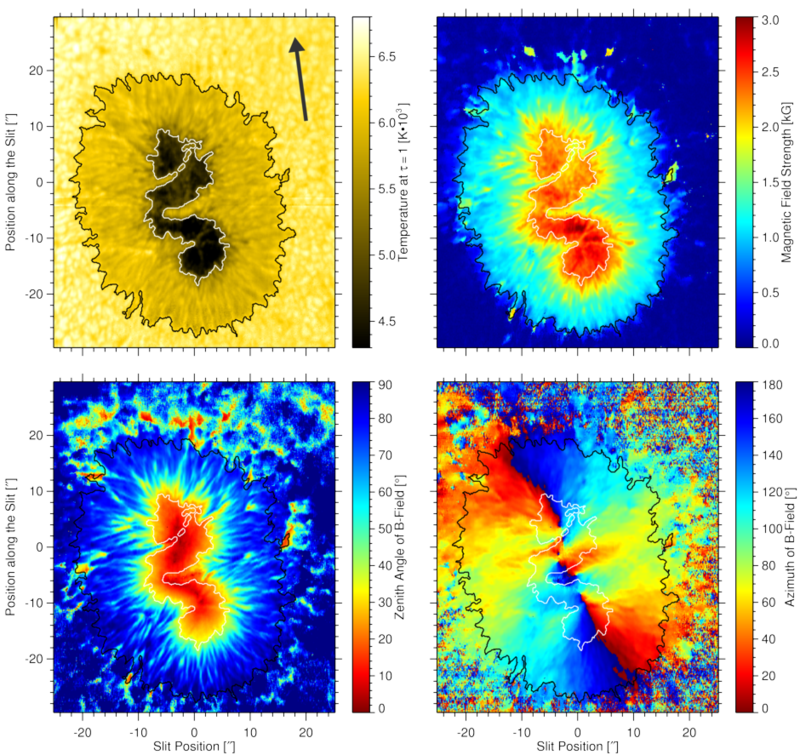}
     \caption{Inversion of a spectropolarimetric raster scan of a
sunspot in AR 12049, observed at 1.568 $\mu$m with GRIS at GREGOR in May
2014. Clockwise from top left: temperature, magnetic field strength,
azimuth and inclination of the magnetic field. Courtesy of GREGOR under
the leadership of the Kiepenheuer-Institut f\"ur Sonnenphysik (KIS) with
the German partners AIP, IAG and MPS, the Spanish IAC and with
contributions by the Czech ASCR.}
         \label{gregor}
   \end{figure*}

\subsection{GREGOR}

{\bf Design and Instrumentation. } The 1.5~m GREGOR telescope on Tenerife, Spain, is the currently largest European solar telescope
\citep[e.g.,][and further papers in the special issue of AN Vol.~333, Issue 9]{gregor2012an}.
Its design started in 1998/1999 and the commissioning in 2012. The early science phase is ongoing from 2014-2015 with access mostly limited to the consortium,
which is led by the German Kiepenheuer Institute. The first open call for proposals from SOLARNET was released in March 2015, but is restricted to EU
and associated countries.

The design consists of a Gregory telescope with three on-axis mirrors and an image derotator further down in the image path,  which is planned to be installed in November 2015.
The large primary mirror suffered from several fabrication problems with the originally planned silicon carbide (Cesic) material and is now made from
light-weighted Zerodur with active cooling. High-order adaptive optics, which allow diffraction-limited observations at 0.08\arcsec\ at 500 nm for good
seeing ($r_0 \ge 10$) are integrated into the path. Three (later four) post focus instruments allow to observe a very large spectral range
from 350 nm to the near Infrared (several micron) and have a field-of-view of up to 150\arcsec. The instruments include a broadband imager
(no polarimetry), the GREGOR Fabry Perot Interferometer
\citep[GFPI,][]{gregorgfpi2013},
a dual collimated Fabry Perot system, and the GREGOR Infrared
Spectrograph
\citep[GRIS,][]{gris2012AN},
which can be combined with the ZIMPOL system
\citep{gregorzimpol2014}.
A stellar spectrograph is planned to be installed in the future.

{\bf Science. }The main science goals include high-resolution photospheric observations
to investigate the structure and dynamics of sunspots and of
granulation, small-scale magnetic field studies to investigate the
presence of a local dynamo, and the largely unexplored chromospheric
magnetic field. Fig~\ref{gregor} shows some first results of the fine
structure of magnetic features obtained from an inversion of GRIS data
of a sunspot with lightbridges (AR 12049). For another example, see Lagg et al. (this issue). GRIS has already shown a good performance in the infrared.

{\bf Future.}  Proposals for the first open observing season in 2016 will be solicited at the end of 2015. It is also planned to replace the M2 mirror, currently made of Cesic, by a version made of Zerodur in the next couple of years, which is expected to improve the contrast and thus the use of the AO. Polarimetric capabilities are a possible upgrade for the stellar spectrograph.

\subsection{SST}

\begin{figure*} 
  \centering
   \includegraphics[width=.7\textwidth]{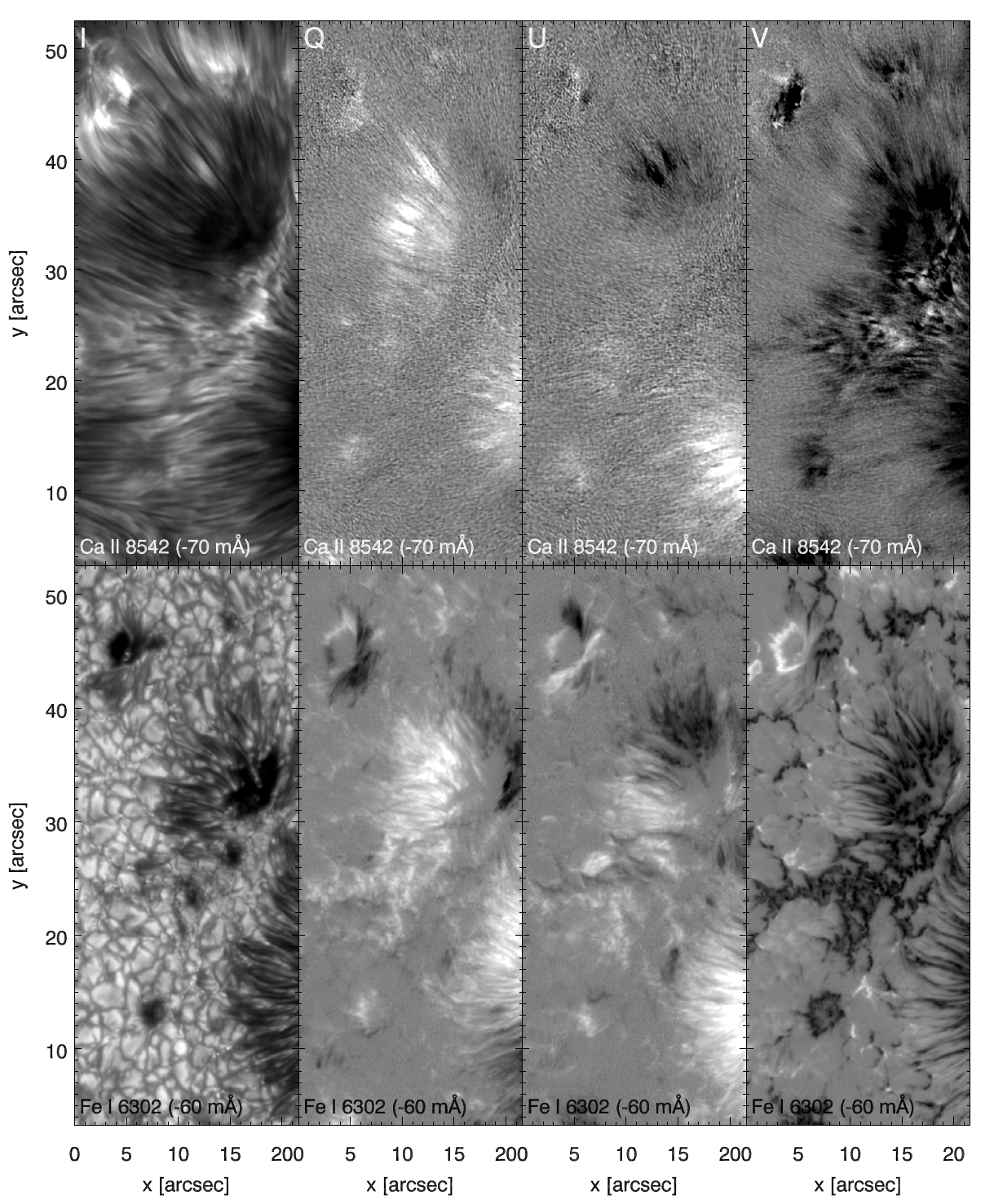}
   \caption{AR11793 taken on 2013-07-19, 13:34UT in the photosphere (bottom row) and chromosphere (top row) in intensity ($I$), linear polarization states ($Q, U$) and circular polarization ($V$), taken with SST/CRISP showing the very high spatial resolution that can be obtained. Courtesy of Jaime de la Cruz Rodr\'iguez (Institute for Solar Physics, Stockholm University).}
        \label{figsst}
  \end{figure*}

{\bf Design and Instrumentation. } The Swedish 1-m Solar Telescope
\citep[SST,][]{scharmeretal2003}
 on La Palma, Spain, has produced some of the highest-resolution magnetograms to date, both in the chromosphere and in the photosphere. Its design consists of a fused silica lens with a clear aperture of 97 cm, a turret with two 1.4 m flat mirrors, and an evacuated tube to minimize air turbulence caused by heating. Chromatic aberration induced by the front lens is corrected by the Schupmann system, which consists of a negative lens (being passed twice by the light) and a mirror, and creates an achromatic image at the secondary focus. With only 6 mirrors in the beam before the optical table, the SST's throughput is very high. An Echelle-Littrow spectrograph (called TRIPPEL) is available for spectroscopy, simultaneously for three spectral windows with a resolution of R$\sim$200000
\citep{kiselmanetal2011}.
For polarimetry, SST's main
 instrument is CRISP, a telecentric dual Fabry Perot system with a spectral resolution of $\sim$60 m\AA\ at 6302~\AA, and an image scale of 0.059\arcsec\ pixel$^{-1}$
\citep{scharmeretal2008,jaime2015}.
The incoming beam is split and the short wavelengths are sent to separate fast cameras, while CRISP records the red part of the light. The polarization modulation was recently upgraded to ferroelectric liquid crystals, which
 change modulation state in less than 1 ms. The modulation is thus limited by the camera exposure time and readout,
giving a speed of $\sim$28 ms per state, i.e.\, 112 ms for a full polarization cycle. After the prefilter for the Fabry Perot, a part of
the light is split off to a separate camera, serving as a broad-band reference for image reconstruction. The remainder of the light passes
through the two Fabry Perot interferometers and the two orthogonal beams after the polarizing beamsplitter are recorded with two separate
cameras. The SST group has pioneered the use of the multi-object multi-frame blind deconvolution (MOMFBD) technique
\citep{vannoortetal2005},
which is applied on the frames already before the demodulation. Coupled with exceptionally good seeing, it produces high spatial resolution, as illustrated in Fig.~\ref{figsst}.

{\bf Science. } Recent scientific results with polarimetry include the discovery of opposite polarities in the penumbra
\citep{scharmeretal2013}
and a study of the effects of umbral flashes and running penumbral waves on the chromospheric magnetic field through inversions of the chromospheric
Ca~II 8542 \AA\ line
\citep{jaimewaves2013}.
 Chromospheric inversions were also used to demonstrate that fibrils often, but not always trace the magnetic field lines
\citep{jaimehector2011}.

{\bf Future. }The next planned upgrade at the SST is the CHROMIS instrument, a Fabry Perot Interferometer to observe the Ca~II H (3934 \AA) and K (3968 \AA) lines.
While its first version is planned without polarimetry, an upgrade would be possible later and it would be interesting to investigate the rather strong scattering polarization of these lines. It would be possible to upgrade TRIPPEL for polarimetry, enabling for example rasters of He 10830 to derive chromospheric magnetic fields, but there are no immediate plans.

\subsection{THEMIS}
{\bf Design and Instrumentation.} The THEMIS telescope on Tenerife, Spain, is a 90 cm Ritchey-Chr\'etien reflector with an alt-az mount. Belonging to the French Centre National de la Recherche Scientifique (CNRS), THEMIS was operated by France \& Italy until 2009, and by France since then, but with a lower level of funding. Its specialty is being virtually polarization-free due to its symmetric design and the polarimeter being placed next to the primary focus, enabling high-accuracy spectropolarimetry. Seeing in and around the telescope is minimized by a helium-filled tube and a specially constructed dome whose small opening coincides and co-rotates with the telescope aperture. A highly complex system of transfer optics and vertically oriented spectrographs feeds light to the instruments. Its design with many optical elements however leads to light loss, especially towards the blue wavelengths. Long-slit multi-line observations are possible with the MTR (MulTi-Raies) instrument \citep{arturo2000mtr}, which can observe up to 6 different wavelength regions, recorded by 12 cameras in a dual-beam setup. THEMIS pioneered several new instrument concepts: The MSDP
\citep[Multi Channel Subtractive Double Pass,][]{mein2002},
allowed to observe two spectral lines simultaneously, with several windows per spectral line and two orthogonal polarization states recorded simultaneously through an elaborate double-pass through a grating. It is no longer available at THEMIS, but a new version is operating at the Meudon Observatory. TUNIS (Tunable Universal Narrowband Imaging Spectrograph) \citep{arturo2010tunis,arturo2011tunis} was developed after MSDP's idea and expanded through a so-called Hadamard mask, encoding full spectral information in the images. After 63 measurements shifting the Hadamard mask to predefined positions, the cube of $x$, $y$, $\lambda$ can be reconstructed, making it a fast instrument compared to regular scanning spectrographs, for example.

{\bf Science.} Due to its low instrumental polarization, THEMIS is well-suited for high-precision polarimetric studies. Studies have focused on magnetic fields in prominences \citep[e.g.,][]{arturoetal2006,schmiederetal2014} and the second solar spectrum \citep{faurobertarnaud2003,faurobertetal2009}.
\citet{lopezariste2012mercury} investigated the scattering polarization of Na in the exosphere of Mercury, which could be used in the future to study its magnetic field.

{\bf Future. }There are no observing campaigns in 2015 and 2016 to allow refurbishing THEMIS' full optical path to enhance the ``polarization-free'' feature of the telescope. It is also planned to install an AO system to improve its spatial resolution. THEMIS plans to resume observations in 2017 and stay operational until the EST comes online.

\begin{figure*} 
  \centering
   \includegraphics[width=.8\textwidth]{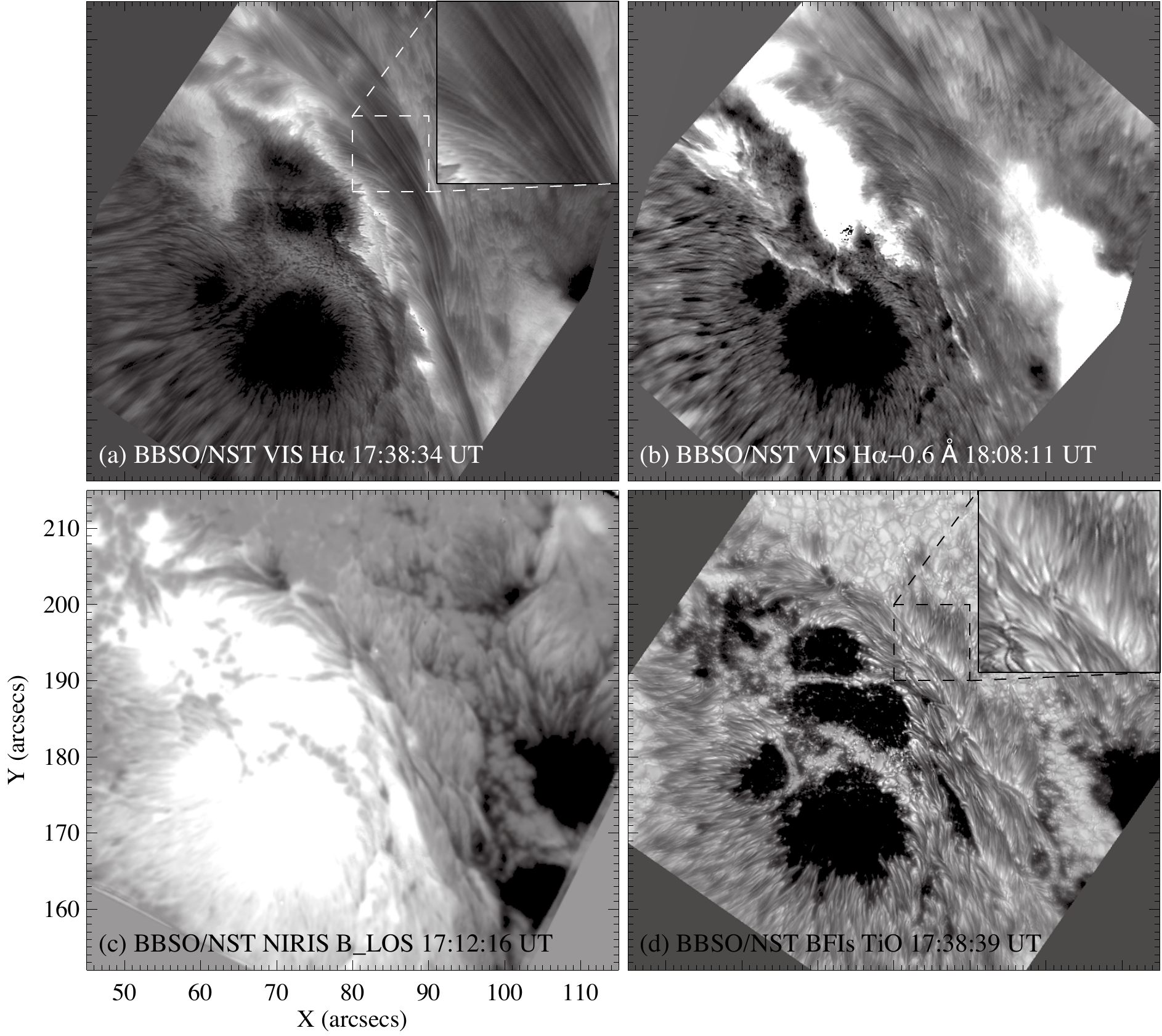}
   \caption{Observation from June 22, 2015 showing NOAA 12371, which produced a long-duration M6.5 flare at 17:39 UT (start) - 18:23 UT (peak) - after midnight (decay). Panels a) and d) show the speckle-reconstructed pre-flare images in H$\alpha$ and TiO, respectively with a plate scale of 0.034\arcsec pixel$^{-1}$. The pre-flare magnetogram from the Fe I 1565 nm line is visible in panel c) with a plate scale of 0.083"/pixel. Panel b) shows a speckle-reconstructed image of the blue wing of H$\alpha$ during the flare peak with two flare ribbons and connecting loops. Courtesy of BBSO, New Jersey Institute of Technology.}
        \label{figniris}
  \end{figure*}

\subsection{BBSO/NST}

{\bf Design and Instrumentation. } The Big Bear Solar Observatory (BBSO) is located in Big Bear, California on a pier inside a lake and is operated by the New
Jersey Institute of Technology. Its location results in less turbulent air motions, which is very beneficial for the seeing.
The 1.6 m New Solar Telescope (NST) was inaugurated in 2009
\citep{goodeetal2010, caoetal2010, goodecao2012},
with the high order adaptive
system following in 2013
\citep{varsikbbso2014}.
It is currently the world's largest solar telescope and has provided some of the highest
resolution images to date. Currently, there are no calls for proposals, but some data and overviews are available on their
website http://www.bbso.njit.edu/ and they plan to accept observing proposals in the near future. NST's design is an off-axis Gregorian system with a wavelength range of 390 nm - 5 $\mu$m. Its instrumentation consists of fast imaging (e.g. Broadband Filter Imagers - BFI - for G-band and TiO and the Visible Imaging Spectrometer - VIS - for H$\alpha$ and Na D2), spectrographs (e.g. Fast Imaging Solar Spectrograph - FISS - used for He 10830 \AA\ and Ca II 8542 \AA\ and the Cryogenic Infrared Spectrograph - Cyra - for observations in the infrared 1 - 5 $\mu$m), and polarimetry. Polarimetric observations are carried out with the Near InfraRed Imaging
Spectropolarimeter
\citep[NIRIS,][]{bbsoniris2012}.
NIRIS is a dual Fabry Perot system in a telecentric configuration
with an 85\arcsec\ FOV with a wavelength coverage from 1.0 to 1.7 microns. As of 2015, the instrument is in routine operations and acquires full-Stokes images in the near infrared line pair at Fe I 1564.85 nm and 1565.29 nm

{\bf Science. } During the past few years, NST studies have focused on fast imaging (no polarimetry yet in the following science highlights). For example, they studied rather rare three-ribbon flares
\citep{wangetal2014},
sunspot oscillations
\citep{yurchyshynetal2015},
flux emergence coupled with a jet
\citep{zengetal2013, vargasetal2014},
the eruption of a flux rope
\citep{wangetal2015},
and the connection between small-scale events in the photosphere and subsequent coronal emission
\citep{jietal2012}.
From observations of a C-flare,
\citet{zengetal2014}
concluded that the He 1083 nm triplet is formed primarily by photoionization of chromospheric plasma followed by radiative recombination. A first result from NIRIS is shown in Fig.~\ref{figniris}.

{\bf Future. } Spectroscopy and spectropolarimetry in the He I 1083.0 nm are the next goal of NIRIS to investigate chromospheric magnetic fields in high resolution. Benefiting from an existing Lyot filter, the NST explored the scientific potential of such observations. First-light for this NIRIS upgrade is expected in 2016. VIS is also being upgraded to dual-FPIs with a diameter of 100 mm each to provide spectroscopic/ polarimetric measurements of the lines in a wavelength range of 550-860 nm.

\subsection{NSO/DST}

The National Solar Observatory (NSO) operates several telescopes, one of
them being the 76~cm Dunn Solar Telescope (DST) at Sacramento Peak in
New Mexico \citep[e.g.][]{zirker1998}.
Inaugurated in 1969, it was the world's leading telescope for many years.

{\bf Design and Instrumentation. } The design of the DST is very
peculiar, with a turret on top of a 41.5~m tower and a 250-ton optical
system suspended on a liquid mercury bearing to counteract the solar
image rotation caused by the alt-az mount.

With its long history, instruments at the DST were constantly upgraded
and are now some of the most versatile in the world. For polarimetric
measurements, the Diffraction Limited Spectropolarimeter
\citep[DLSP,][]{sankarasubramanian2004},
the Interferometric Bidimensional Spectropolarimeter
\citep[IBIS,][]{cavallini2006},
the Facility
Infrared Spectropolarimeter
\citep[FIRS,][]{jaegglietal2010},
or the Spectro-Polarimeter for
Infrared and Optical Regions
\citep[SPINOR,][]{socasnavarroetal2006}
can be used, covering the range
from visible to infrared and from spectroscopy to imaging. IBIS is
currently one of the few instruments capable of  imaging polarimetry in the
chromospheric Ca~II 8542 \AA\ line. The ASP and DLSP helped in the
development of the highly successful SP instrument onboard Hinode. The
DST led the development of solar adaptive optics, and was the first
telescope to be equipped with high-order adaptive optics
\citep{rimmeleetal2004},
enabling diffraction limited studies of small-scale features.

{\bf Science. } The DST enabled many discoveries, whose list would exceed the
scope of this paper. Some examples are sunspot and penumbral
oscillations
\citep{beckersschultz1972},
the first study of the subsurface structure of sunspots
\citep{thomasetal1982},
the confirmation that the penumbral magnetic field is composed of two
components using the ASP
\citep{litesetal1993},
the discovery of
small-scale short-lived ($\sim$minutes) horizontal magnetic fields in
the internetwork
\citep{litesetal1996}
 and the first magnetic map of a prominence \citep{casinietal2003}. Using the multi-line capabilities of SPINOR, the magnetic field of the quiet Sun \citep{socasnavarroetal2008}, of spicules \citep{socasnavarroelmore2005} and its 3D structure in sunspots \citep{socasnavarro2005} could be studied. FIRS allowed to confirm through spectropolarimetric observations in the He 10830 \AA\ line that the superpenumbral fibrils trace the magnetic field \citep{schadetal2013}.
More recently, polarimetric observations by IBIS allowed to study the inclination change of the magnetic field during the formation of a penumbra \citep{romanoetal2014}, and a
coordinated observing campaign with Hinode and IRIS was led from the DST, resulting in the
``best-observed X-flare''
\citep{kleintetal2015}.
In recent years, the DST carried out 3 cycles of service mode
observations. While common in night-time astronomy and for
satellite missions, they were a first in solar ground-based observations
and a preparation for significantly more efficient DKIST observing modes.

{\bf Future. }NSO plans to cease operations
of the DST by the end of 2017 while preparing for first light with their
new 4-m DKIST on Maui, but negotiations with a consortium of
universities and institutes that could operate the facility after 2017
are ongoing.

\subsection{NLST}

The 2~m National Large Solar Telescope (NLST) is a proposed project in India
\citep{hasan2012}.
The design studies and site selection are complete, but the construction is not funded (yet).
The proposal is awaiting formal clearance from the government of India. The fabrication will take about 3.5 years and it is expected to take place
from late 2016 - 2020. The project is led by the Indian Institute of Astrophysics.

{\bf Design and Instrumentation. } The NLST's Gregorian design is similar to the GREGOR telescope, but with fewer mirrors, leading to a higher throughput.
It is designed to observe from 380-2500~nm. The planned first-light instruments include a narrow band imager and a spectropolarimeter,
whose design has not been finalized yet. After four years of site surveys, Hanle near the border to Tibet was selected as primary site.
At an altitude of 4500~m, it has very low water vapor and favorable weather.

Currently, India is commissioning the 50 cm off-axis MAST telescope.
The integration of its 19 actuator AO and of two LiNbO$_3$ Fabry Perot interferometers to observe the photospheric 6173 \AA\ line and
the chromospheric 8542 \AA\ line is ongoing
\citep{bayannaetal2014}.
The post-focus instruments include broad band and tunable Fabry-Perot narrow band imaging instruments;
a high resolution spectropolarimeter and an echelle spectrograph for night time astronomy.

\subsection{NVST}

The 98.5 cm New Vacuum Solar Telescope (NVST) at Fuxian Lake is a Chinese telescope, which has recently been upgraded for high-resolution
observations
\citep{yangetal2014, liuetal2014}.

{\bf Design and Instrumentation. }  As a vacuum telescope, it contains an entrance window, which is followed by a
modified Gregorian design. It is capable of observing from 300 to 2500 nm.
A new high-order adaptive optics system with 151 actuators was developed in 2015 and
great care is taken to reduce the local seeing, with the location of the telescope (lakeshore),
by cooling the building's roof with a shallow pool of water, with a wind screen, and with the vacuum
inside the telescope tube. There are two optical paths after the AO, one leading to three broadband
channels (H-$\alpha$, TiO and G-band), which are imaged by three separate cameras, and the other to a
spectrograph system, which is being upgraded to polarimetric capabilities. A polarimeter for observations
of the Fe~I 5324 \AA\  line and the line pair at 5247--5250 \AA\ has been developed and calibrations are
ongoing and a near-IR magnetograph for the 1.56 $\mu$m line is planned.

{\bf Future. }An ambitious Chinese project is the Chinese Giant Solar Telescope, currently in the planning phase with no final design yet
\citep{liuetal2012,liuetal2014cgst}.
Current ideas include a ring of mirrors on a 8 m diameter aperture, which would have the light-gathering power equivalent to a 5-m telescope.
The telescope is planned to work mostly in the infrared (0.3 -- 15 $\mu$m),
because the seeing variations are lower. One goal is to measure the magnetic field in lines near 10 $\mu$m, e.g. 10.4 $\mu$m.
After 5 years of site surveys, two candidate sites have been selected (one lake and one mountain site), with no final decision expected in the near future.

\subsection{EST}

The European Solar Telescope (EST) is a joint project of several European institutions
with the goal of building a 4 m telescope on La Palma in the Canary Islands
\citep{est2013}.
Its conceptual design study was finished successfully in 2011 and it is currently in the detailed design stage (2013-2017).
The construction phase might take place earliest from 2018-2023, but its funding, estimated at 150 M\euro, is not approved yet.

{\bf Design and Instrumentation. } EST's design consists of an on-axis Gregorian on an alt-az mount, similar to GREGOR.
The instruments are not defined yet, but will include a broad band imager, a narrow-band
tunable imager and a grating spectrograph. The advantage of the EST compared to the DKIST
is that the design is favorable for polarimetric observations (though at the expense of coronal observations)
and one of EST's requirements is high-resolution spectropolarimetry simultaneously in the photosphere and in the
chromosphere. The telescope Mueller matrix is unity for all wavelengths and independent of elevation and azimuth
and the calibration optics are located on axis, which is the ideal configuration for polarimetry.
Another feature of EST is to include multi-conjugate adaptive optics (MCAO) in the beam, which so
far has not been done before, even though first MCAO tests are ongoing. Distortions of the incoming
wavefront through atmospheric turbulence occur at multiple heights. Current AO systems
include one deformable mirror, which allows for a good correction within the so-called isoplanatic
patch. In other words, at its edges, away from the AO lockpoint, the FOV is more variable and possibly
blurry. MCAO consists of multiple deformable mirrors and would allow to correct for turbulence occurring
at multiple heights, giving a more uniform FOV. Because the EST is on nearly the opposite side of the world
than the DKIST, a near-continuous coverage of a target would be possible.

\subsection{DKIST}

\begin{figure*}
   \centering
\includegraphics[width=.4\textwidth]{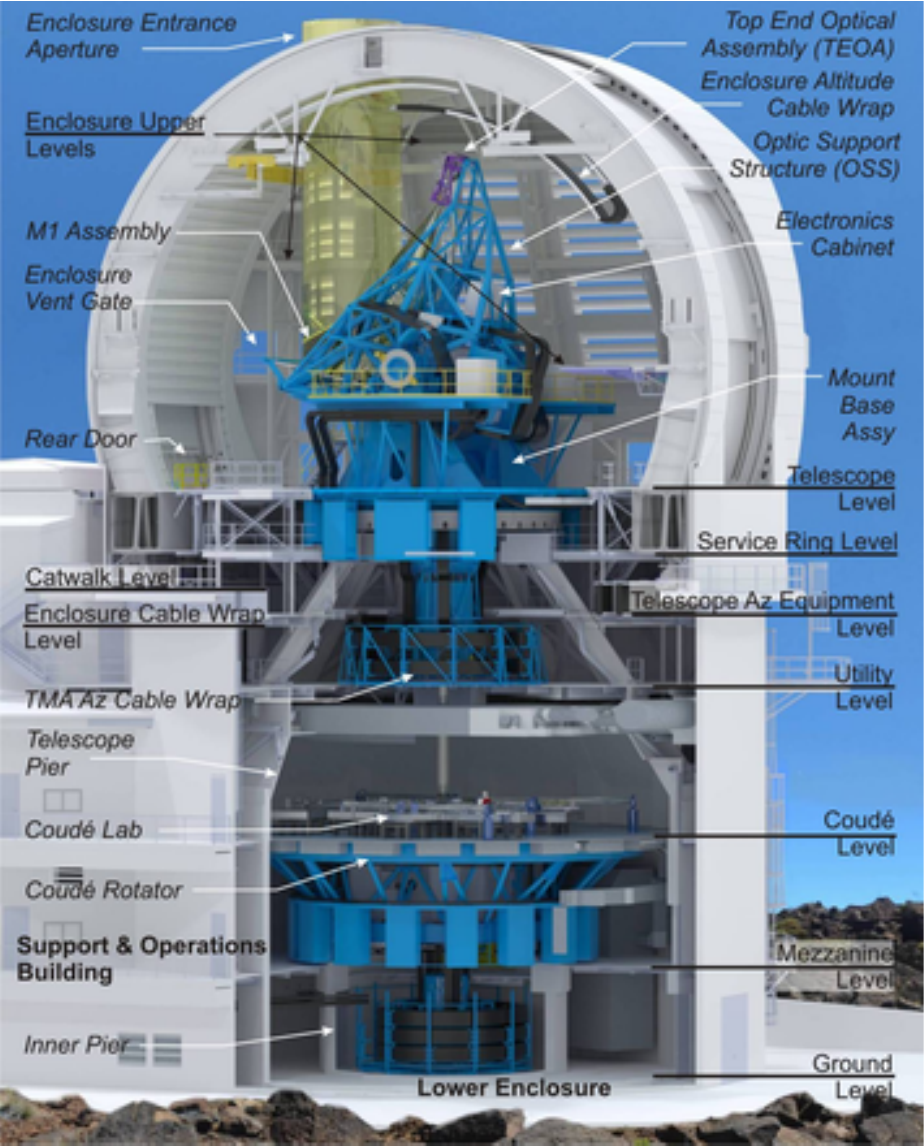}
\includegraphics[width=.55\textwidth]{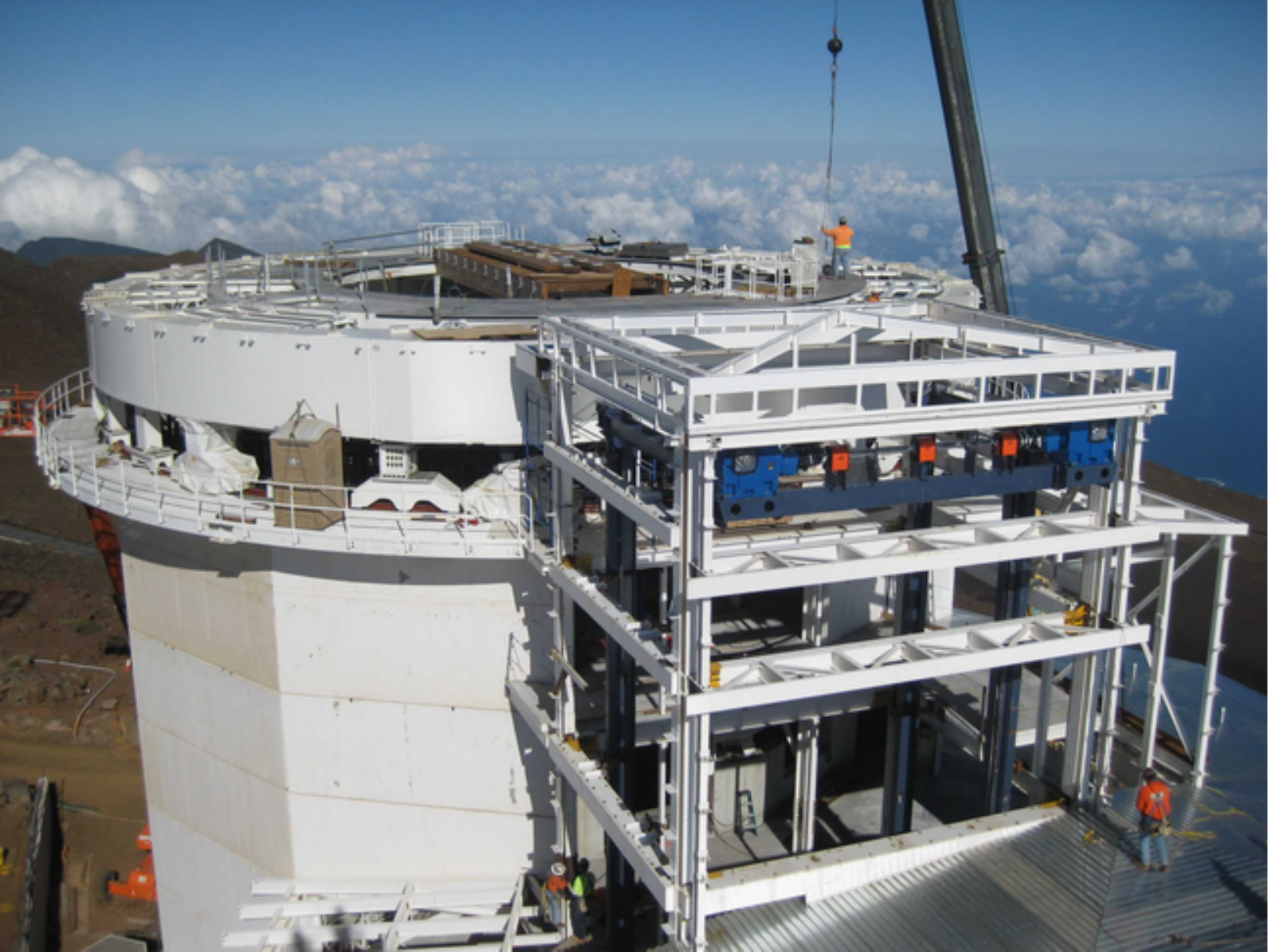}
    \caption{Left: DKIST telescope enclosure with details labeled.
Right: DKIST construction progress on the summit of Haleakal\={a}
(Hawaii), April 30, 2015. Courtesy of NSO/DKIST.}
         \label{dkist}
   \end{figure*}

The 4~m Daniel K. Inouye Solar Telescope (DKIST, formerly ATST) led by NSO will be the world's largest
solar telescope when it is commissioned in 2019. The four meter aperture is a factor of more than 6 in light
gathering power compared to the currently largest telescope, the NST. With a wavelength range of 380 - 5000~nm,
and possibly up to 28000~nm (28 $\mu$m) for second generation instruments, the DKIST will provide unprecedented
magnetic field measurements from the photosphere to the corona up to 1.5 solar radii.

{\bf Design and Instrumentation. }  The DKIST is currently under construction on Haleakala on Maui HI. The enclosure, which was shipped from Spain
to Hawaii, will be installed in 2015. Its design allows for thermal control and dust mitigation, especially
important for coronal observations where any scattering through dust on the mirrors needs to be avoided. The
Telescope Mount Assembly will be installed in 2016-2017 and the optics system installation and integration is
planned for 2017-2018. After the instrument commissioning, first light is planned for the middle of 2019. The
off-axis Gregory setup was chosen in view of the low straylight for coronal observations. However, the off-axis
design results in significant polarization, variable over the day, and will need to be calibrated carefully.

There will be five first-generation instruments, four of which are capable of polarimetric measurements.
Contrary to many other telescopes, most of them can be operated simultaneously (with the exception of the
coronal instrument Cryo-NIRSP) by splitting spectral ranges off the incoming beam. The Visible
Spectropolarimeter (VISP), led by the High Altitude Observatory, consists of an Echelle spectrograph
from 380 - 900 nm, and will record three spectral ranges on different cameras simultaneously. The
Visible Tunable Filter (VTF), led by the Kiepenheuer Institute, is a dual Fabry Perot with coatings
that permit observations from 550 - 860~nm. The Diffraction Limited Near Infrared Spectro-Polarimeter
(DL-NIRSP), led by the University of Hawaii, consists of a fiber-fed multi-slit spectrograph and a
selection of cameras that cover the range of 900 - 2300~nm. Additionally to these polarimetric
instruments, the Visible Broadband Imager (VBI), led by NSO, will record the intensity in selected
filters from 390-860 nm and the images will be speckle-reconstructed directly after their acquisition
to improve the spatial resolution. The fourth polarimetric instrument is the Cryogenic Near Infrared
Spectro-Polarimeter (Cryo-NIRSP), led by the University of Hawaii, which is the only instrument that
does not allow beam-sharing and which will not utilize the adaptive optics. Its wavelength range 1000-5000~nm
is optimized for diffraction-limited coronal observations. Polarimetry is possible up to 4000~nm. For a more
complete overview of the instruments, see
\citet{Elmoreetal2014}.

{\bf Science. }The DKIST will be the largest advance in ground-based solar observations in several decades.
It is expected that its high spatial and temporal resolution and its polarimetric capabilities
will lead to ground-breaking discoveries, especially for chromospheric magnetic fields, turbulent
magnetic fields and local dynamo mechanisms, coronal magnetism, and the photospheric fine-structure,
including that of sunspots.

\section{Next Generation Instrumentation}

\begin{figure*} 
  \centering
   \includegraphics[width=.95\textwidth]{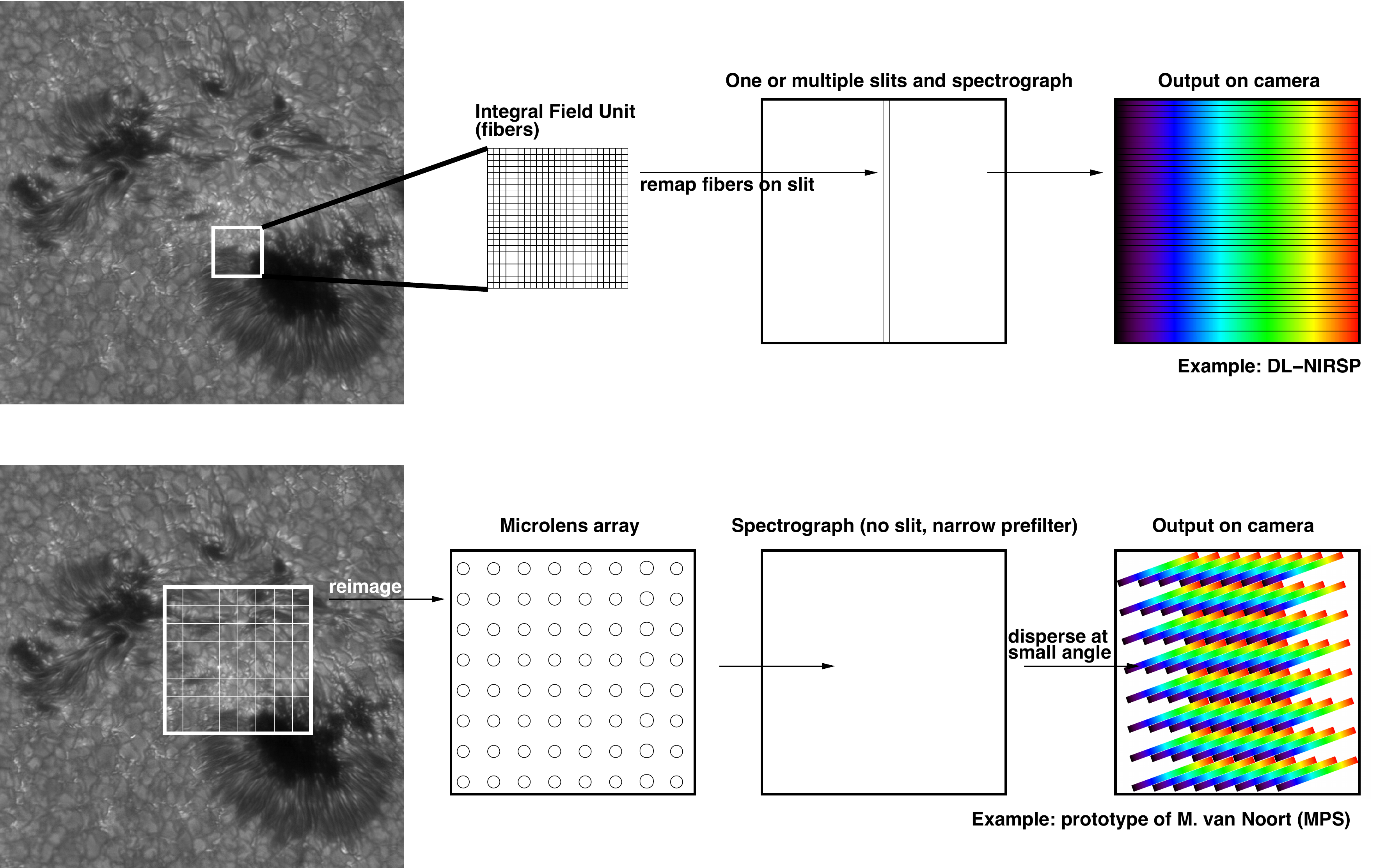}
   \caption{Two principles of how to record $[x,y,\lambda]$ simultaneously. Both are currently being used in prototype instruments. Figure adapted from M. van Noorts and T. Schads versions.}
        \label{xylambda}
  \end{figure*}

Even with the significant advances of DKIST, some desired observations will still not be possible.
For example, Hanle imaging to investigate turbulent magnetic fields, or extremely fast flare polarimetry are outside the realm of
any currently planned DKIST instrument. While a 4 m telescope will provide the necessary light gathering power
and the spatial resolution, possibly with MCAO, it would be desirable to record the cube $[x,y,\lambda,Stokes]$ quasi-simultaneously,
or at least faster than any seeing-variation ($\sim$hundred Hz). This requires a sophisticated instrument design, plus very fast
detectors that can modulate and read-out faster than $\sim$100 Hz.

\subsection{Spatial and spectral data simultaneously}

For the instrument design, there are several possibilities to obtain $[x,y,\lambda]$ simultaneously,
with two examples depicted in Fig.~\ref{xylambda}. One option is a fiber array (integral field unit) that subdivides the
focal plane into parts that are then fed into a spectrograph and re-arranged onto the detector. An example for this type
of instrument is the DL-NIRSP of DKIST, which is planned to have a mode with 19200 fibers that are fed into 5 slits.
A slight disadvantage is the rather limited FOV (for good spatial resolution) and thus a requirement to scan to obtain a picture of e.g. a full sunspot.

Another option is to re-image the focal plane with a
microlens array, in principle shrinking the pixels. They
are then dispersed by a spectrograph at a small rotation
angle, so that they can fit on a detector. A prefilter
needs to be used to avoid an overlap of spectra of different image
elements. FOV and spectral range can be
traded for another. A prototype of such an instrument is currently being tested
by M. van Noort at the Max Planck Institute for Solar
System Research.

\begin{figure*}[ht]
\centerline{\includegraphics[width=0.85\textwidth,clip=]{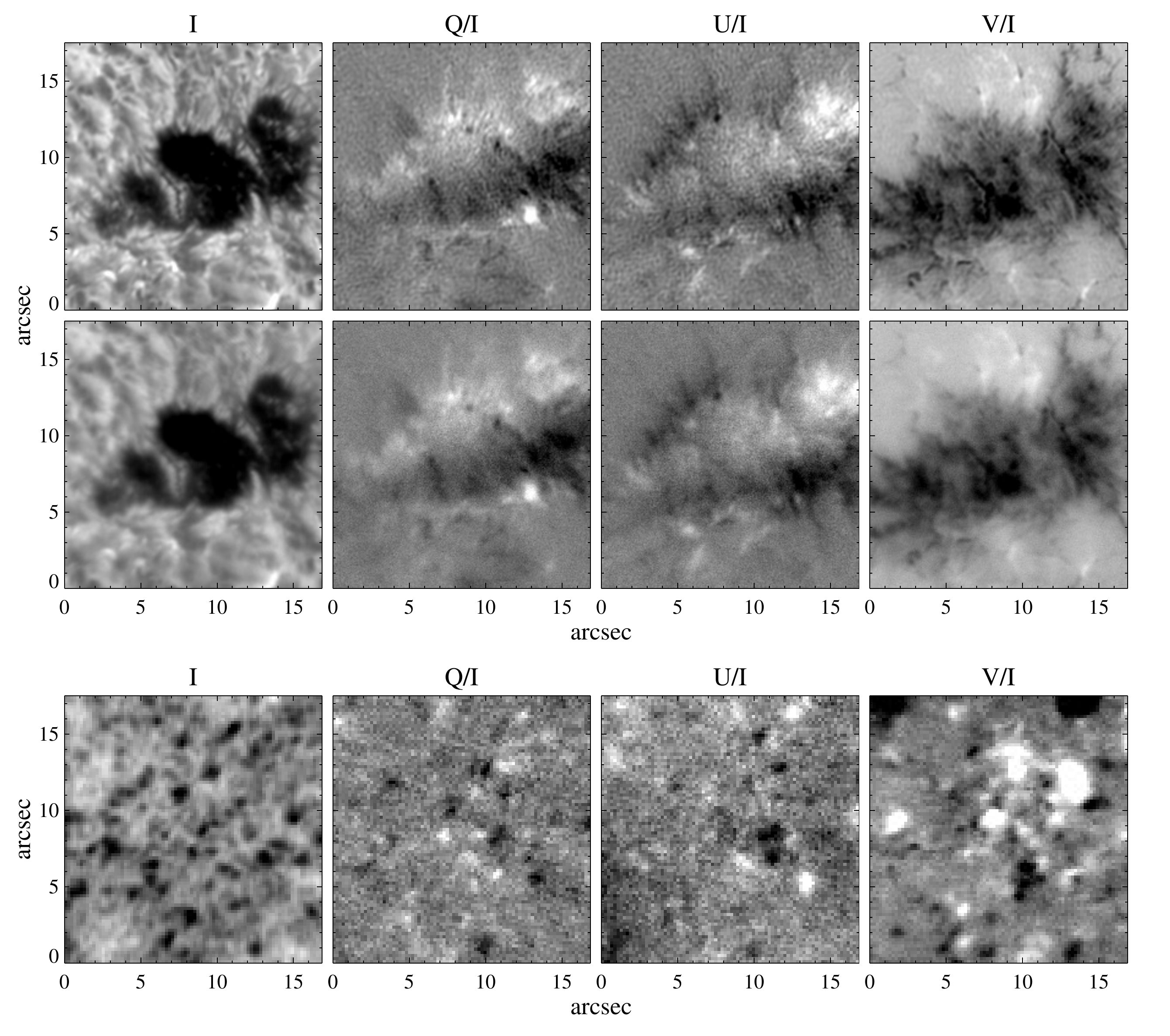}}
\caption{Example results obtained with the FSP prototype at the TESOS filtergraph on the VTT.
Upper panel: Part of a small active region recorded in Fe I 630.2~nm. The four columns represent the full state of polarization, i.e. Stokes $I, Q/I, U/I$, and $V/I$ with grayscales $I: <I>\pm40\%, Q/I, U/I: \pm 8\%, V/I \pm 15\%$.
Each image of the upper row is a result of a MOMFBD restoration involving 640 individual frames, spanning 1.6~s. For comparison, the images in the second row have just been averaged, which
significantly reduces spatial resolution.
Lower panel: Small-scale and very faint magnetic fields of the quiet Sun, recorded in the same Fe I line, with a very high polarimetric sensitivity of 0.02\%. A 3x3 pixels binning has been applied to further increase S/N. Grayscales: $I: <I>\pm 10\%, Q/I, U/I: \pm 0.25\%, V/I: \pm 0.5\%$. Courtesy A. Feller. }
\label{fsp}
\end{figure*}

A third option would be to employ an image slicer that separates the FOV into equal slices that are re-imaged by
mirrors onto a spectrograph. Within one of these slices, the spatial information is retained, which is not the case
for principles a) and b). Such an option is currently being explored for the EST
\citep{calcinesetal2013,calcinesetal2014}.

\subsection{Fast Solar Polarimeter}
A very promising project for extremely fast modulation and readout is the \textit{Fast Solar Polarimeter (FSP)},
led by A. Feller at the Max Planck Institute for Solar System Research
\citep{felleretal2014}.

The FSP overcomes several drawbacks of ZIMPOL, while reaching similar polarimetric sensitivity.
ZIMPOL's pixels are not square (due to the masked rows used for charge shifting), and rather large, which is not ideal for high-resolution images.
The overall ZIMPOL throughput is also relatively low, even with the microlenses that focus light from the masked rows to the unmasked rows.
Also, ZIMPOL's overhead is rather large (above 100\%) for short exposures, which is not ideal for fast changing solar features, such as flares.
The FSP tackles these problems by a different design, and still reaches frame rates high enough to sufficiently suppress seeing-induced crosstalk. It has a
specially designed pnCCD sensor, which can read out at speeds up to 400 fps. This is achieved by shifting the charges of each half
of the CCD to one side (split frame transfer), and by parallel readout of all sensor columns. The polarization modulation is achieved with two ferro-electric liquid crystals,
which operate at the same frame rates, reaching a duty cycle of $>95\%$. A polarimetric accuracy of $10^{-4} I{_c}$ at sub-arcsec
resolution can be achieved by summing images below solar evolution time scales ($\sim$ minutes). Alternatively,
about 1 restored set of Stokes images  can be obtained per second with a typical S/N of order 300-500 by combining some 400 single frames.
This allows to apply image restoration with a cadence down to about 1~s, which to our knowledge is not possible so far with any other ground-based solar
polarimeter.
The first prototype of the FSP with a 256x256 pixel pnCCD was tested at the VTT (see Fig.~\ref{fsp}). Currently, a 1024x1024 pnCCD is in development,
which will be used in a dual beam setup. First observations are planned for 2016.
The FSP may be the instrument, which will allow to clarify the question of the inconsistent field strengths derived from scattering
polarization observations (cf Section~\ref{lgap}), or to enable fast polarimetry during flares.

\section{Future space initiatives and projects with relevance to solar magnetometry}
In this section we present a brief overview of space missions with relevance
to solar magnetometry. We restrict the list to space missions, which have not yet
been launched, but rather are in the preparation or even planning phase. Details of some of these missions are also mentioned in A. Lagg et al. (this issue).

\subsection{CLASP - Chromosheric Lyman-alpha SpectroPolarimeter}

The observation and interpretation of linear polarization caused by
scattering in the solar atmosphere, and its modification
by magnetic fields via the Hanle effect, has been identified as a diagnostic
tool for solar magnetic fields already early
\citep[see][]{stenflostenholm1976}.
From a technological as well as from interpretational aspects, the field of scattering polarimetry
has been boosted by the advent of highly sensitive polarimeters, mainly ZIMPOL in its various evolutionary generations.
 Ground based observations of the second solar spectrum, both with ZIMPOL and at the THEMIS telescope, stimulated the
development of new theoretical concepts and numerical tools, which brought us
into a situation, where we can finally use scattering polarization as a
diagnostic tool for solar magnetometry.

The first attempts to measure the scattering polarization in the Lyman alpha
line of hydrogen and in this way retrieve information on the physical
conditions in the upper chromosphere and transition region was undertaken by
Stenflo and collaborators in the late 1970s. A small polarimeter for
measuring linear polarization around 121~nm
\citep{stenfloetal1976}
was installed on the russian
Intercosmos 16 satellite. Although the experiment failed due to contamination
issues, it was able to set an upper limit to the polarization degree, however with a
large uncertainty
\citep{stenfloetal1980}.
\citet{stenfloetal1980}
claim
that "The average polarization of the Lyman alpha solar limb was found to be less than 1\%.
It is indicated that future improved VUV polarization
measurements may be a diagnostic tool for chromospheric and coronal magnetic
fields and for the three-dimensional geometry of the emitting structures."

Although the essential usefulness of UV polarimetry is widely accepted, and the development of UV polarization technology is explicitly
recognized in
the ASTRONET Infrastructure Roadmap as a "perceived gap" (see p. 73 of the Astronet Infrastructure Roadmap 2008), until today there has been no further instrument
operating in this wavelength regime.
Lyman alpha polarimeters have been
proposed twice as strawman instruments for ESA missions, first for the
COMPASS mission proposed in response to the call for the M2 slot
\citep{fineschietal2007},
later as part of
the SOLMEX mission proposed in 2008 in response to ESAs M3 call
\citep{peteretal2012}.
It is obvious that - given the explorative character of such an instrument -
a demonstration of the principle including the successful interpretation of the harvested data is
mandatory.

To this end, by initiative of the Japanese Space Agency JAXA, together with NASA and with European contributions,
the CLASP mission was established. CLASP stands for Chromospheric Lyman-alpha
SpectroPolarimeter and is an explorative sounding rocket experiment with the aim of
demonstrating the capabilities of UV spectropolarimetry from space
\citep[c.f.][and references therein]{kanoetal2012,kuboetal2014}.
CLASP consists of a 279~mm diameter Cassegrain telescope feeding a dual-beam
spectrograph equipped with a rotating half-wave plate modulator for the
measurement of Stokes I, Q, and U in and close to the Lyman alpha line in a
wavelength range from 121.1nm - 122.1nm.

The target is the chromosphere and transition region as seen on disk
radially from the solar limb. The spatial resolution is 1.5\arcsec\ (set by the slit
width) times 2.9\arcsec\ (along the slit) in a field of view of 1.5\arcsec\ times
400\arcsec.
With a spectral resolution of 0.01nm, in 5 min observing time, a noise level
of 0.1\% can be achieved when further binning the data by averaging along the slit. A context imager with an
angular resolution of 2.2\arcsec\ images a field of view of 550\arcsec
$\times$ 550\arcsec\ in order to identify the structures, which are observed by the
spectropolarimeter.

The requirements on polarimetric sensitivity and accuracy are derived from
models of the expected polarization signals
 \citep[c.f.][and references therein]{ishikawaetal2014a},
based on reported estimates of
the field strength in chromospheric structures
\citep{trujillobuenoetal2011}.
For 5-50 G, the calculated
polarization profiles indicate that a sensitivity of 0.1\% - 0.2\% is
necessary. While in the line core a sensitivity of 0.1\% in the linear polarization is aimed for, the
sensitivity in the wings of the line can be decreased to 0.5\% due to the
larger polarization signals there.
 More recent 3-D simulations of scattering polarization and the Hanle effect in MHD chromospheric
models have been undertaken by
\citet{stepan2015} and \citet{stepanetal2015}.

The spectropolarimeter employs a spherical constant-line-spaced grating with 3000 grooves per mm and
uses the plus and minus first diffraction orders for dual beam polarimetry.
Two camera mirrors feed two cameras.
Two reflective polarization analyzers are employed in front of the cameras. This setup maximizes the photon efficiency
and allows for high sensitivity polarimetry. To this end, all polarimetric error sources must be controlled
by a thorough calibration concept and a rigorous polarimetric error budget tracking
\citep{ishikawaetal2014}.
The linear polarization is modulated by a rotating wave plate made from MgF$_2$ with
a  retardance of half a wave, rotating at a period of 4.8 seconds. With
sixteen exposures per round, the optimum exposure time of 0.3 seconds per
frame is achieved.

In order to cope with the demanding false-light suppression requirements and for thermal
reasons, the primary mirror acts as a cold mirror.  The coating of this mirror has high reflectivity only around
121~nm (54~\% as measured on test samples),  but is transparent in the visible, thus transmitting most of the incoming solar energy to a thermally insulated light trap.

The international effort is led by Japan, which provides the experiment  structure,  telescope  optics,
polarimeter  optics,  slitjaw  optics,  and the waveplate
rotation mechanism. The sounding  rocket,  flight  operations,  CCD
cameras, and avionics are under responsibility of NASA.
Spain contributes by modeling of the Hanle effect, while
the modeling tools for the chromosphere are developed under Norwegian responsibility. France
contributes to the mission by providing the diffraction grating.

At the time of writing, CLASP was scheduled for launch in September 2015. During the manuscript review process, CLASP has been launched, but no results have been announced yet. The
total mission observing time was 5~min.
If the mission is successful, CLASP will pave the way for the application of UV polarimetry from space.

\subsection{Sunrise}

Although not being real space missions according to classical definition,
stratospheric balloons offer a valid alternative for a low cost access to a
near space like environment
\citep[for a recent review see][and references therein]{gaskinetal2014}.
Since the technological aspects of balloon
borne observatories are much closer to those of space instruments than to those
of ground-based facilities, we list Sunrise here.

Sunrise is the largest and most complex UV/Vis solar observatory up to date, which could
escape the disturbing influences of the terrestrial atmosphere. Based on a
1-m aperture optical telescope and two scientific post focus instruments in
the near UV and the visible range of the solar spectrum, it was designed for
highest resolution observations of solar surface magnetic fields.
Sunrise is an international mission led by the Max-Planck-Institute of Solar
System Research (MPS) in G\"{o}ttingen, Germany, together with the German Kiepenheuer
Institute of solar physics in Freiburg, a Spanish consortium under leadership
of the IAC (until 2012) and the IAA (since 2013), and the High
Altitude Observatory in Boulder, USA. Also involved is the Lockheed Martin Solar and
Astrophysics laboratory in Palo Alto. The ballooning aspects are under
responsibility of NASA's Columbia Ballooning Facility.

In two 6-day flights from ESRANGE space center near Kiruna (Sweden) to Northern
Canada in Summer 2009 and in summer 2013, respectively, Sunrise could demonstrate its unique potential.  Despite of some technical issues in both flights, the magnetograms recorded of the extremely quiet Sun
in 2009 and of active regions in 2013 represent some of the highest resolution seeing free magnetic
field and Doppler maps of the photosphere. The co-spatial brightness maps of
the photosphere and low chromosphere in the near UV (down to 220nm, which is not accessible from the ground) are the highest
contrast images of the solar surface ever recorded
\citep{hirzbergeretal2010},
thanks to the absence of
seeing and by exploiting the contrast transfer capabilities of a 1-m
aperture
telescope in the near UV. In this wavelength range, the sensitivity of the
brightness to temperature fluctuations is very pronounced because of the
steepness of the Planck curve; this helped in mapping the temperature
structure and thus identifying magnetic bright points with high sensitivity
\citep{riethmuelleretal2010},
while the magnetic field and the flows were
directly measured in the visible. A number of scientific insights into the physics of small-scale solar
surface magnetism resulted from these observations
\citep[see e.g.][and references therein]{solankietal2011}.

Sunrise contains an instrument suite, which consists of a near UV
filtergraph
\citep[SuFI,][]{gandorferetal2011}
and an imaging magnetograph
\citep[IMaX,][]{valentinetal2011},
which are both fed with light from a 1~m aperture diameter Gregory telescope
 by a light distribution unit
 \citep[ISLiD,][]{gandorferetal2011}.
 This
unit also provides the high resolution images for a correlation tracker and wave front sensor
\citep[CWS,][]{berkefeldetal2011},
which reduces residual pointing errors
of the telescope by a fast tip tilt mirror inside the light distribution
unit, and keeps control of the telescope alignment.
The IMaX magnetograph is based on two key technologies:  wavelength
selection is done by a tunable solid state etalon, and the polarization
analysis makes use of two nematic liquid crystals. Both technologies have
proven to be applicable in near-space conditions and represent a big step
forward in our efforts to build compact and lightweight polarimeters for
space applications. The Sunrise experience thus directly influences the
development of the {\it Polarimetric and Helioseismic Imager} onboard {\it Solar
Orbiter}
(see below) and is considered a necessary and successful predecessor.

Ballooning offers unique opportunities for testing new instrumental concepts
for their usage in space. Most solar instruments have reached a technological
complexity, which cannot be transferred to space in one step. This
applies in particular to new detector architectures and the complex
electronics systems associated with them. Special newly developed sensors like
the one of the Fast Solar Polarimeter are today in a prototype status for
ground-based demonstration of the concept. It is only natural to expand these
efforts to use such sensors in Sunrise, and maybe - after successful
qualification- some day in a new space observatory. Even without being
affected by seeing, very high resolution polarimeters in space suffer from
the internal vibrations of the spacecraft, which cannot be completely
avoided. This becomes more and more the dominant cost driver in the development of
satellite platforms for solar observations. Fast detectors greatly relax
the requirements on the residual pointing errors and thus will some day enable polarimetry
at extremely high resolution from space. The jitter spectrum of a
stratospheric balloon is typically much more demanding and thus represents a
 worst case validation. Therefore, if Sunrise will fly again,  it would be desirable to have a
 prototype of a near-space FSP-type sensor on board.

\subsection{Solar Orbiter and its polarimetric and helioseismic imager
PHI}

More than twenty years after the launch of SoHO, a new European solar mission
 will be launched to space in 2018, Solar
Orbiter, a collaborative mission of ESA and NASA.

Solar Orbiter is conceived and designed to clarify the magnetic coupling
from the photosphere throughout the solar atmosphere into the inner heliosphere,
and aims at providing answers to questions like the following: How and where do the solar wind plasma
and magnetic field originate in the corona?
How do solar transients drive heliospheric
variability?
How do solar eruptions produce energetic
particle radiation that fills the heliosphere?
 How does the solar dynamo work and
drive the connections between the Sun
and the heliosphere?

Carrying a suite of remote sensing instruments and an in-situ analysis
package, Solar Orbiter is an integrated and unique approach to heliophysics, since it
combines aspects of a space solar observatory with the characteristics of an encounter mission.
Solar Orbiter is not particularly
focussed on magnetometry, but the mission will offer unique opportunities
to study surface fields and to probe the solar dynamo
\citep{loeptienetal2014}.

Solar Orbiter's
science can be addressed thanks to a unique orbit design:
During a 3.5 year transfer orbit the spacecraft undergoes several gravity
assist maneuvers (GAM) at Venus and Earth, which help the spacecraft to lose
orbital energy, and thus allow {\it Solar Orbiter} to come close to the Sun. After
the
second GAM at Venus, {\it Solar Orbiter} begins its operational phase. From then on
its orbit is in a three-to-two resonance with Venus, such that after each third orbit
the inclination of the orbital plane with respect to the ecliptical plane can
be increased by a further gravity assist. This particular feature
gives {\it Solar Orbiter} access to the high latitude regions of the Sun for
the first time.

While the in-situ instrument suite will be operational over the full orbit,
the remote sensing instruments will be used in three distinct science phases
per orbit, the perihelion passage, and the phases of maximum and minimum
solar latitude.
During the perihelion passages {\it Solar Orbiter} can follow the evolution of surface
structures and solar features not only from close-by, but in addition under
practically unchanged geometrical viewing conditions for several days, thanks
to a corotating vantage point.

The {\it Solar Orbiter} Instrumentation can be grouped in three major packages,
each consisting of several instruments:

\begin{itemize}
\item Field Package:
Radio and Plasma Wave Analyzer and Magnetometer.
\item Particle Package:
Energetic Particle Detector and Solar Wind Plasma
Analyser
\item Solar remote sensing instrumentation:
Visible-light Imager and Magnetograph, Extreme Ultraviolet Spectrometer, EUV Imager, Coronagraph, and
Spectrometer/Telescope for Imaging X-rays,
Heliospheric Imager.
\end{itemize}

From all instruments onboard, the visible light imager and magnetograph,
called "Polarimetric and Helioseismic Imager PHI" is of highest relevance for
solar surface magnetometry. Together with the other high resolution instruments it observes the same target region on the solar surface with an identical angular
sampling of 0.5~arcsec per pixel.  By combining its observations with the other instruments on Solar Orbiter, PHI will address the magnetic coupling between the different atmospheric layers.
Extrapolations of the magnetic field observed
by PHI into the Sun's upper atmosphere and heliosphere will provide the
information needed for other optical and in-situ instruments to analyze and
understand the data recorded by them in a proper physical context.

\begin{figure*}[ht]
\centerline{\includegraphics[width=1.0\textwidth,clip=]{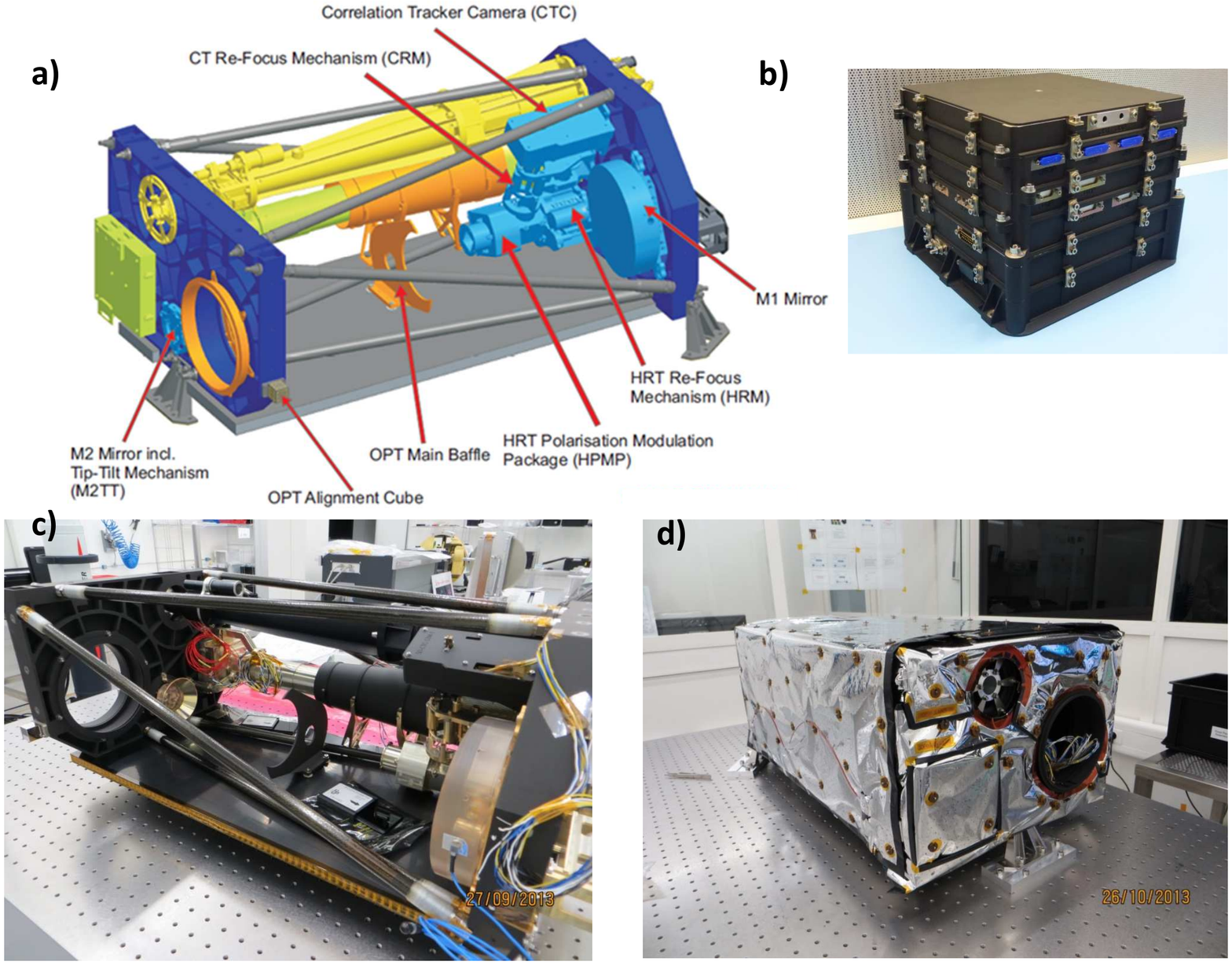}}
\caption{Images of the PHI instrument:
a) CAD representation of the PHI Optics Unit with FDT tube (yellow) and HRT subsystems;
b) structural thermal model (STM) of the electronics unit;
c) STM of the optics unit without multi-layer insulation (MLI);
d) same, but now with mounted MLI. From \citet{solankietal2015a}} 
\label{block}
\end{figure*}

 The instruments on Solar Orbiter are very challenging to build.
No space magnetograph has ever flown in such
a difficult environment, with the spacecraft following a strongly elliptical
trajectory, leading to significant thermal changes in the course of an orbit.
The technical description below follows closely the more complete instrument descriptions by
\citet{gandorferetal2011}
and
\citet{solankietal2015a}.
For the sake of completeness and readability, we
present a shortened version here.
PHI makes use of the Doppler- and Zeeman-effects in a
single spectral line of neutral iron at 617.3~nm. The physical information
 is decoded from two-dimensional intensity maps at
 six wavelength points within this line, while
 four polarization states at each wavelength point are measured.
In order to obtain the abovementioned observables, PHI is a
 diffraction limited,
 wavelength tunable,
 quasi-monochromatic,
 polarization sensitive
imager with two telescopes, which (alternatively) feed a common filtergraph and focal plane
array (for illustration see Fig.~\ref{block}):
 The High Resolution Telescope (HRT) provides a
restricted FOV of 16.8~arcmin squared and achieves a
spatial resolution that, near the closest perihelion
pass, will correspond to about 200~km on the Sun.
It is designed as a decentered Ritchey-Chr\'{e}tien telescope with a
 pupil of 140~mm diameter.
The all-refractive Full Disk Telescope (FDT), with a FOV of 2.1$^\circ$
in diameter and a pixel size corresponding to 730~km (at 0.28~AU),
provides a complete view of the full solar disk during all
orbital phases.
These two telescopes are used alternatively and their
selection is made by a feed selection mechanism.

Both telescopes are protected from the intense solar flux by special
heat-rejecting entrance windows, which are part of the heat-shield assembly
of the spacecraft. These multilayer filters have more than 80\% \ transmittance
in a narrow notch around
the science wavelength, while effectively blocking the remaining parts of the spectrum from 200~nm to
the far infrared by reflection. Only a fraction of the total energy is absorbed in the window, which
acts a a passive thermal element by emitting part of the thermal radiation to
cold space; emission of infrared
radiation into the instrument cavity is minimized by a low emissivity
coating on the backside of the window
(acting at the same time as an anti-reflection coating for the science wavelength). Thus the
heat load into the instruments can be substantially reduced, while
preserving the high photometric and polarimetric accuracy of PHI.

The filtergraph unit FG uses two key technologies with heritage from IMaX
onboard Sunrise:
A LiNbO$_3$ solid state etalon in a telecentric configuration selects a passband of
100~ m\AA \ width. Applying a high voltage across the crystal allows changing the
refractive index of the material and its thickness, and thus tuning the passband in wavelength
across the spectral line.
\citet{gensemerfarrant2014}
report on the fabrication
technology for these etalons,
whose   absolute   thickness,   approximately   250~$\mu$m,   is
controlled  to  $<$10~nm,  maintaining  a  thickness  uniformity  of  $<$1~nm  over  the  60~mm  aperture.
A 3~\AA \ wide prefilter acts as an order sorter for the
Fabry-Perot channel spectrum.
The polarimetric analysis is performed by two Polarization Modulation
Packages (PMP) in each of the telescopes.
Each PMP consists of two nematic liquid crystal retarders, followed by a linear polarizer as an analyzer.
The liquid crystal variable retarders have been successfully qualified for
the use in Solar Orbiter
\citep{alvarezherreroetal2011}.
Both, the FG and the PMPs, are thermally insulated from the Optics Unit and
actively temperature stabilized.

The opto-mechanical arrangement is designed to
operate
in a wide temperature range. To this end the optics unit
  structure consists of a combination of AlBeMet (an
 aluminum-beryllium alloy) and low expansion carbon-fibre reinforced plastic.
An internal image stabilization system in the HRT channel system acts on the active secondary
mirror, which
greatly reduces residual pointing error by the spacecraft to levels compatible
with high resolution polarimetry.

The error signal for the piezo-driven mirror support is derived from a
correlation tracker camera inside the HRT. For details on the Image
stabilization system we refer to
\citet{carmonaetal2014}.
The focal plane assembly is built around a 2048 by 2048 pixel Active
Pixel Sensor (APS), which is specially designed and manufactured for this
instrument. It delivers 10 frames per second, which are read out in
synchronism with the switching of the polarization modulators.

Besides all the technological complexity of the cutting-edge hardware, the most critical
aspect of the instrument lies, however, in the data reduction strategy:
The extremely limited telemetry rate and the large amount of scientific information
retrieved from the PHI instrument demand a sophisticated on-board data
reduction and necessitate to employ a non-linear, least-square,
inversion technique, which numerically solves the radiative transfer
equation on board.  This inversion is based on the Milne-Eddington approximation and is provided by the highly adaptable onboard s/w, two powerful
reprogrammable FPGAs
\citep{fietheetal2012},
and a large non-volatile data storage unit.

\subsection{Solar-C}

Solar-C
\citep{watanabe2014}
 is an international vision for the ultimate future solar space observatory,
initiated by Japan in the wake of the highly successful Hinode mission.

Solar-C is a JAXA led solar space observatory with the overarching science
goal of "understanding the Sun\'\ s magnetized atmosphere from bottom to top by

- understanding the dynamic structuring and mass and energy loading of the
solar atmosphere

- understanding the basic plasma processes at work throughout the solar
atmosphere

- understand the causes of solar activity, which affects our natural and
technical environment" \citep{solankietal2015prop}.

To this end, Solar-C represents an extremely powerful space observatory in a
geosynchronous orbit, with a payload that significantly improves our
capabilities in imaging and spectropolarimetry in the UV, visible, and near
infrared with respect to what is available today or foreseen in the near
future. The mission concept and strawman instrumentation has been conceived and designed
by the efforts of the ISAS/JAXA Solar-C working group, and a summary of the mission has been
published by
\citet{watanabe2014}.
A strong European Contribution to Solar-C is envisaged. A first proposal to ESA
in response to the call for the M4 mission
\citep{solankietal2015prop}
was very highly ranked but was not selected in this round.

Solar-C aims at probing the different temperature regimes in the solar
atmosphere with unprecedented angular resolution and at simultaneously covering the
entire solar atmosphere from the solar surface to the outer corona. Three
telescopes, working in distinct spectral regimes, are required for this task:

-SUVIT, a 1.4~m aperture UV/VIS/NIR telescope for imaging and spectropolarimetry with an order of
magnitude increase in photon collecting area over the largest solar
telescope in space. With its instrumentation suite, which will be described
below, this telescopes covers seamlessly the solar photosphere and
chromosphere.

- EUVST, a 30~cm aperture VUV telescope feeding a spectrometer for imaging spectroscopy
covering line formation temperatures ranging from the chromosphere to the
hottest parts of the corona. Over this extended temperature range, EUVST will
have a resolution and effective collecting area an order of magnitude better
than available today.  EUVST is based on the LEMUR instrument described by
\citet{teriacaetal2012}.

- HCI, a next generation high resolution extreme-ultraviolet imager with 32~cm aperture
operating at high cadence in multiple spectral lines sampling the
chromosphere and corona.

The telescope, which is of most relevance to magnetometry, is the Solar
Ultraviolet and Visible Telescope SUVIT
\citep{suematsuetal2014solarc}.
SUVIT will for the first time
measure chromospheric magnetic fields from space and is expected to obtain photospheric
vector fields at very high resolution with the highest sensitivity ever for subarcsec resolution. SUVIT is a telescope with an aperture of 1.4~m, providing an angular
resolution of 0.1\arcsec\ in combination with spectropolarimetry in wavelengths
ranging from 280~nm to 1080~nm.

With its aperture, SUVIT would have at least 10 times higher sensitivity than
any other current or planned space mission. The increase in spatial resolving
power and light gathering power directly results in an order of magnitude
improvement in sensitivity of magnetic flux, electric current and magnetic
and kinetic energy.
Another major advantage of SUVIT is that it samples a large range of line
formation heights almost seamlessly.
In order to achieve this, the telescope feeds several highly dedicated
post-focus instruments via a common optical interface unit. This interface
unit provides each instrument with a collimated beam with the required
spectral bands; it also contains the common tip-tilt correction mirror for
stabilizing image motion caused by the residual pointing error of the S/C.
A similar strategy - although for a much narrower spectral range - has been successfully implemented in the SOT of
Hinode
\citep{suematsuetal2008}.
The post-focus instrumentation
consists of a tunable filtergraph (FG) with
polarimetric sensitivity for the spectral range from 525~nm to 1083~nm. The baseline design employs
a Lyot filter equipped with liquid crystals as variable retardation plates, which allows tuning in
wavelength without any moving or rotating parts.
The classical echelle-type grating spectropolarimeter (SP) covers the same
wavelength range and also possesses vector magnetometric capabilities in a
variety of spectral lines. In order to achieve the key science goals of
retrieving the magnetic field information over the full atmospheric height,
nearly simultaneous observations in several lines are mandatory. This can be
achieved in two different ways: Either the grating is moved and the spectral
bands are recorded in series with a single camera, or - in order to obtain
strictly simultaneous recordings - three cameras are employed; in both options
interference filters act as order sorters for the echelle grating. While the
three camera solution - to first sight - seems more complex, it should not be
forgotten, that it spares two mechanisms, which is always beneficial in terms
of reliability and microvibration control. The final decision on which option to
follow must be done by a careful assessment during the study phase of the
mission.
Photospheric and chromospheric magnetograms with simultaneous two-dimensional
coverage can be achieved only by an integral field unit (IFU). Such a unit is
foreseen for the SP as well, it maps a 9" squared patch of the
field of view onto three entrance slits of the SP, located parallel to the
"normal" entrance slit. It can be based on an array of optical fibers (Lin
2012) or on micro image slicers
\citep{suematsuetal2014slicer}.

The capabilities of high-resolution imaging in the near UV for providing new insights into the photosphere and chromosphere
have been
established by the SUFI imager onboard the Sunrise balloon-borne stratospheric observatory.
IRIS has unveiled the upper chromosphere in the light of the Mg II h\& k
lines, uncovering the dynamic and energetic phenomena by high resolution UV
spectroscopy.
In the wake of these achievements, the Ultraviolet Imager and Spectropolarimeter (UBIS) will further advance those observations and
provide UV filtergrams at much higher spatial resolution.
In addition, expanding the capabilities of IRIS, UBIS aims at
spectropolarimetry in the Mg II h \& k lines, allowing for magnetometry in
the upper chromosphere.
To this end, UBIS consists of two sub-units, a re-imager with filter wheels,
and an optional spectropolarimeter behind it. The spectropolarimeter should be considered in this stage as an optional instrument, since the potential of
spectropolarimetry in the Mg II lines still needs to be thoroughly
demonstrated. Recent significant progress in forward modelling suggests that the chromospheric magnetic field can be reliably diagnosed
with these lines
\citep{belluzzijavier2012}.

\section{Outlook}

We are at an exciting stage in solar magnetic measurements and the next major missions and telescopes are within reach. By 2020, we expect our newest ground- and space-based telescopes, DKIST and Solar Orbiter, to provide unprecedented data.

The 4~m ground-based DKIST, currently in construction and scheduled for first light in 2019, will improve our current spatial resolution (diffraction limit) by a factor of 2.5. This will for example allow to compare features of sizes of fractions of arcseconds in sunspots, such as umbral dots, or penumbral grains to the most advanced simulations to investigate details of sunspot dynamics, formation and decay. The polarimetric sensitivity of better than 10$^{-4} I_c$, coupled with the large aperture will hopefully allow us to solve the mystery of the orientation of small-scale magnetic fields in the quiet Sun and thus to investigate the local dynamo processes. Employing fast polarimetry will enable us to study the magnetic structure of rapidly changing features, such as flares, filament eruptions, or dynamic flows.

After Solar Orbiter's launch, currently scheduled for 2018, and its cruise phase, we will for the first time have a polarimeter on a significantly inclined orbit around the Sun allowing us to study its polar regions. We will not only be able to study the mechanism of flux transport towards the poles in more detail, but also combine the observations with a whole suite of 9 additional instruments on Solar Orbiter, giving us a more complete picture of the different solar layers.

Plans for the far future include the EST, a 4~m class ground-based telescope in Europe, with a design specifically optimized for polarimetric measurements
and Solar-C, the successor of the Japanese Hinode satellite, that will hopefully enable us to study chromospheric polarimetry systematically and from space for the first time, shedding light on a highly dynamic and yet not well explored solar layer.

\section*{Acknowledgments}
We thank the experts of the telescopes and instruments for their advice and information, in particular, Wenda Cao, Gianna Cauzzi, Manolo Collados, Jaime de la Cruz
Rodriguez, Alex Feller, Bernard Gelly,
S S Hasan, Bruce Lites, Chang Liu, Arturo Lopez Ariste, Zhong Liu, Valentin Martinez Pillet, Rolf Schlichenmaier, Alexandra Tritschler,  Michiel van Noort, and Haimin Wang.

\bibliographystyle{aa}  
\bibliography{journals,papers}

\begin{thebibliography}{118}
\expandafter\ifx\csname natexlab\endcsname\relax\def\natexlab#1{#1}\fi

\bibitem[{Alvarez-Herrero {et~al.}(2011)Alvarez-Herrero, Uribe-Patarroyo,
  García~Parejo, Vargas, Heredero, Restrepo, Martínez-Pillet, del
  Toro~Iniesta, López, Fineschi, Capobianco, Georges, López, Boer, \&
  Manolis}]{alvarezherreroetal2011}
Alvarez-Herrero, A., Uribe-Patarroyo, N., García~Parejo, P., {et~al.} 2011, in
  Proc. SPIE, Vol. 8160, 81600Y--81600Y--18

\bibitem[{{Aschwanden}(2011)}]{aschwanden2011}
{Aschwanden}, M.~J. 2011, Living Reviews in Solar Physics, 8, 5

\bibitem[{{Asensio Ramos}(2009)}]{andres2009}
{Asensio Ramos}, A. 2009, \apj, 701, 1032

\bibitem[{{Barthol} {et~al.}(2011){Barthol}, {Gandorfer}, {Solanki},
  {Sch{\"u}ssler}, {Chares}, {Curdt}, {Deutsch}, {Feller}, {Germerott},
  {Grauf}, {Heerlein}, {Hirzberger}, {Kolleck}, {Meller}, {M{\"u}ller},
  {Riethm{\"u}ller}, {Tomasch}, {Kn{\"o}lker}, {Lites}, {Card}, {Elmore},
  {Fox}, {Lecinski}, {Nelson}, {Summers}, {Watt}, {Mart{\'{\i}}nez Pillet},
  {Bonet}, {Schmidt}, {Berkefeld}, {Title}, {Domingo}, {Gasent Blesa}, {Del
  Toro Iniesta}, {L{\'o}pez Jim{\'e}nez}, {{\'A}lvarez-Herrero},
  {Sabau-Graziati}, {Widani}, {Haberler}, {H{\"a}rtel}, {Kampf}, {Levin},
  {P{\'e}rez Grande}, {Sanz-Andr{\'e}s}, \& {Schmidt}}]{bartholetal2011}
{Barthol}, P., {Gandorfer}, A., {Solanki}, S.~K., {et~al.} 2011, \solphys, 268,
  1

\bibitem[{{Beckers} \& {Schultz}(1972)}]{beckersschultz1972}
{Beckers}, J.~M. \& {Schultz}, R.~B. 1972, \solphys, 27, 61

\bibitem[{{Belluzzi} \& {Trujillo Bueno}(2012)}]{belluzzijavier2012}
{Belluzzi}, L. \& {Trujillo Bueno}, J. 2012, \apjl, 750, L11

\bibitem[{{Berkefeld} {et~al.}(2011){Berkefeld}, {Schmidt}, {Soltau}, {Bell},
  {Doerr}, {Feger}, {Friedlein}, {Gerber}, {Heidecke}, {Kentischer},
  {v.~D.~L{\"u}he}, {Sigwarth}, {W{\"a}lde}, {Barthol}, {Deutsch}, {Gandorfer},
  {Germerott}, {Grauf}, {Meller}, {{\'A}lvarez-Herrero}, {Kn{\"o}lker},
  {Mart{\'{\i}}nez Pillet}, {Solanki}, \& {Title}}]{berkefeldetal2011}
{Berkefeld}, T., {Schmidt}, W., {Soltau}, D., {et~al.} 2011, \solphys, 268, 103

\bibitem[{{Bernasconi} {et~al.}(2000){Bernasconi}, {Rust}, {Eaton}, \&
  {Murphy}}]{bernasconietal2000}
{Bernasconi}, P.~N., {Rust}, D.~M., {Eaton}, H.~A., \& {Murphy}, G.~A. 2000, in
  SPIE Conf. Ser., Vol. 4014, Airborne Telescope Systems, ed. R.~K. {Melugin}
  \& H.-P. {R{\"o}ser}, 214--225

\bibitem[{{Borrero} \& {Kobel}(2011)}]{borrerokobel2011}
{Borrero}, J.~M. \& {Kobel}, P. 2011, \aap, 527, A29

\bibitem[{{Calcines} {et~al.}(2013){Calcines}, {L{\'o}pez}, \&
  {Collados}}]{calcinesetal2013}
{Calcines}, A., {L{\'o}pez}, R.~L., \& {Collados}, M. 2013, Journal of
  Astronomical Instrumentation, 2, 50009

\bibitem[{{Calcines} {et~al.}(2014){Calcines}, {L{\'o}pez}, {Collados}, \&
  {Vega Reyes}}]{calcinesetal2014}
{Calcines}, A., {L{\'o}pez}, R.~L., {Collados}, M., \& {Vega Reyes}, N. 2014,
  in SPIE Conf. Ser., Vol. 9147, 3

\bibitem[{{Cao} {et~al.}(2012){Cao}, {Goode}, {Ahn}, {Gorceix}, {Schmidt}, \&
  {Lin}}]{bbsoniris2012}
{Cao}, W., {Goode}, P.~R., {Ahn}, K., {et~al.} 2012, in Astronomical Society of
  the Pacific Conference Series, Vol. 463, Second ATST-EAST Meeting: Magnetic
  Fields from the Photosphere to the Corona., ed. T.~R. {Rimmele},
  A.~{Tritschler}, F.~{W{\"o}ger}, M.~{Collados Vera}, H.~{Socas-Navarro},
  R.~{Schlichenmaier}, M.~{Carlsson}, T.~{Berger}, A.~{Cadavid}, P.~R.
  {Gilbert}, P.~R. {Goode}, \& M.~{Kn{\"o}lker}, 291

\bibitem[{{Cao} {et~al.}(2010){Cao}, {Gorceix}, {Coulter}, {Ahn}, {Rimmele}, \&
  {Goode}}]{caoetal2010}
{Cao}, W., {Gorceix}, N., {Coulter}, R., {et~al.} 2010, Astronomische
  Nachrichten, 331, 636

\bibitem[{Carmona {et~al.}(2014)Carmona, Gómez, Roma, Casas, López, Bosch,
  Herms, Sabater, Volkmer, Heidecke, Maue, Nakai, \& Schmidt}]{carmonaetal2014}
Carmona, M., Gómez, J.~M., Roma, D., {et~al.} 2014, in Proc. SPIE, Vol. 9150,
  91501U--91501U--15

\bibitem[{{Casini} {et~al.}(2003){Casini}, {L{\'o}pez Ariste}, {Tomczyk}, \&
  {Lites}}]{casinietal2003}
{Casini}, R., {L{\'o}pez Ariste}, A., {Tomczyk}, S., \& {Lites}, B.~W. 2003,
  \apjl, 598, L67

\bibitem[{{Cavallini}(2006)}]{cavallini2006}
{Cavallini}, F. 2006, \solphys, 236, 415

\bibitem[{{Collados} {et~al.}(2013){Collados}, {Bettonvil}, {Cavaller},
  {Ermolli}, {Gelly}, {P{\'e}rez}, {Socas-Navarro}, {Soltau}, {Volkmer}, \&
  {EST Team}}]{est2013}
{Collados}, M., {Bettonvil}, F., {Cavaller}, L., {et~al.} 2013, \memsai, 84,
  379

\bibitem[{{Collados} {et~al.}(2012){Collados}, {L{\'o}pez}, {P{\'a}ez},
  {Hern{\'a}ndez}, {Reyes}, {Calcines}, {Ballesteros}, {D{\'{\i}}az}, {Denker},
  {Lagg}, {Schlichenmaier}, {Schmidt}, {Solanki}, {Strassmeier}, {von der
  L{\"u}he}, \& {Volkmer}}]{gris2012AN}
{Collados}, M., {L{\'o}pez}, R., {P{\'a}ez}, E., {et~al.} 2012, Astronomische
  Nachrichten, 333, 872

\bibitem[{{de la Cruz Rodr{\'{\i}}guez} {et~al.}(2015){de la Cruz
  Rodr{\'{\i}}guez}, {L{\"o}fdahl}, {S{\"u}tterlin}, {Hillberg}, \& {Rouppe van
  der Voort}}]{jaime2015}
{de la Cruz Rodr{\'{\i}}guez}, J., {L{\"o}fdahl}, M.~G., {S{\"u}tterlin}, P.,
  {Hillberg}, T., \& {Rouppe van der Voort}, L. 2015, \aap, 573, A40

\bibitem[{{de la Cruz Rodr{\'{\i}}guez} {et~al.}(2013){de la Cruz
  Rodr{\'{\i}}guez}, {Rouppe van der Voort}, {Socas-Navarro}, \& {van
  Noort}}]{jaimewaves2013}
{de la Cruz Rodr{\'{\i}}guez}, J., {Rouppe van der Voort}, L., {Socas-Navarro},
  H., \& {van Noort}, M. 2013, \aap, 556, A115

\bibitem[{{de la Cruz Rodr{\'{\i}}guez} \&
  {Socas-Navarro}(2011)}]{jaimehector2011}
{de la Cruz Rodr{\'{\i}}guez}, J. \& {Socas-Navarro}, H. 2011, \aap, 527, L8

\bibitem[{{de la Cruz Rodr{\'{\i}}guez} {et~al.}(2012){de la Cruz
  Rodr{\'{\i}}guez}, {Socas-Navarro}, {Carlsson}, \&
  {Leenaarts}}]{jaimeetal2012}
{de la Cruz Rodr{\'{\i}}guez}, J., {Socas-Navarro}, H., {Carlsson}, M., \&
  {Leenaarts}, J. 2012, \aap, 543, A34

\bibitem[{{Elmore} {et~al.}(2014){Elmore}, {Rimmele}, {Casini}, {Hegwer},
  {Kuhn}, {Lin}, {McMullin}, {Reardon}, {Schmidt}, {Tritschler}, \&
  {W{\"o}ger}}]{Elmoreetal2014}
{Elmore}, D.~F., {Rimmele}, T., {Casini}, R., {et~al.} 2014, in SPIE Conf.
  Ser., Vol. 9147, 7

\bibitem[{{Faurobert} \& {Arnaud}(2003)}]{faurobertarnaud2003}
{Faurobert}, M. \& {Arnaud}, J. 2003, \aap, 412, 555

\bibitem[{{Faurobert} {et~al.}(2009){Faurobert}, {Derouich}, {Bommier}, \&
  {Arnaud}}]{faurobertetal2009}
{Faurobert}, M., {Derouich}, M., {Bommier}, V., \& {Arnaud}, J. 2009, \aap,
  493, 201

\bibitem[{{Feller} {et~al.}(2014){Feller}, {Iglesias}, {Nagaraju}, {Solanki},
  \& {Ihle}}]{felleretal2014}
{Feller}, A., {Iglesias}, F.~A., {Nagaraju}, K., {Solanki}, S.~K., \& {Ihle},
  S. 2014, in ASP Conference Series, Vol. 489, Solar Polarization 7, ed. K.~N.
  {Nagendra}, J.~O. {Stenflo}, Q.~{Qu}, \& M.~{Samooprna}, 271

\bibitem[{Fiethe {et~al.}(2012)Fiethe, Bubenhagen, Lange, Michalik, Michel,
  Woch, \& Hirzberger}]{fietheetal2012}
Fiethe, B., Bubenhagen, F., Lange, T., {et~al.} 2012, in Adaptive Hardware and
  Systems (AHS), 2012 NASA/ESA Conference on, 31--37

\bibitem[{{Fineschi} {et~al.}(2007){Fineschi}, {Solanki}, \&
  Team}]{fineschietal2007}
{Fineschi}, S., {Solanki}, S., \& Team, C. 2007, Proposal to ESA Cosmic Vision

\bibitem[{{Gandorfer} {et~al.}(2011){Gandorfer}, {Grauf}, {Barthol},
  {Riethm{\"u}ller}, {Solanki}, {Chares}, {Deutsch}, {Ebert}, {Feller},
  {Germerott}, {Heerlein}, {Heinrichs}, {Hirche}, {Hirzberger}, {Kolleck},
  {Meller}, {M{\"u}ller}, {Sch{\"a}fer}, {Tomasch}, {Kn{\"o}lker},
  {Mart{\'{\i}}nez Pillet}, {Bonet}, {Schmidt}, {Berkefeld}, {Feger},
  {Heidecke}, {Soltau}, {Tischenberg}, {Fischer}, {Title}, {Anwand}, \&
  {Schmidt}}]{gandorferetal2011}
{Gandorfer}, A., {Grauf}, B., {Barthol}, P., {et~al.} 2011, \solphys, 268, 35

\bibitem[{{Gandorfer} \& {Povel}(1997)}]{Gandorferpovel1997}
{Gandorfer}, A.~M. \& {Povel}, H.~P. 1997, \aap, 328, 381

\bibitem[{{Gandorfer} {et~al.}(2004){Gandorfer}, {Steiner}, {Aebersold},
  {Egger}, {Feller}, {Gisler}, {Hagenbuch}, \& {Stenflo}}]{Gandorferetal2004}
{Gandorfer}, A.~M., {Steiner}, H.~P.~P.~P., {Aebersold}, F., {et~al.} 2004,
  \aap, 422, 703

\bibitem[{{Gaskin} {et~al.}(2014){Gaskin}, {Smith}, \&
  {Jones}}]{gaskinetal2014}
{Gaskin}, J.~A., {Smith}, I.~S., \& {Jones}, W.~V. 2014, Journal of
  Astronomical Instrumentation, 3, 3001

\bibitem[{{Gensemer} \& {Farrant}(2014)}]{gensemerfarrant2014}
{Gensemer}, S.~D. \& {Farrant}, D. 2014, Advanced Optical Technologies, 3, 309

\bibitem[{{Goode} \& {Cao}(2012)}]{goodecao2012}
{Goode}, P.~R. \& {Cao}, W. 2012, in ASP Conf. Ser., Vol. 463, Second ATST-EAST
  Meeting: Magnetic Fields from the Photosphere to the Corona., ed. T.~R.
  {Rimmele}, A.~{Tritschler}, F.~{W{\"o}ger}, M.~{Collados Vera},
  H.~{Socas-Navarro}, R.~{Schlichenmaier}, M.~{Carlsson}, T.~{Berger},
  A.~{Cadavid}, P.~R. {Gilbert}, P.~R. {Goode}, \& M.~{Kn{\"o}lker}, 357

\bibitem[{{Goode} {et~al.}(2010){Goode}, {Coulter}, {Gorceix}, {Yurchyshyn}, \&
  {Cao}}]{goodeetal2010}
{Goode}, P.~R., {Coulter}, R., {Gorceix}, N., {Yurchyshyn}, V., \& {Cao}, W.
  2010, Astronomische Nachrichten, 331, 620

\bibitem[{{Harrison} {et~al.}(2009){Harrison}, {Davies}, {Rouillard}, {Davis},
  {Eyles}, {Bewsher}, {Crothers}, {Howard}, {Sheeley}, {Vourlidas}, {Webb},
  {Brown}, \& {Dorrian}}]{harrisonetal2009}
{Harrison}, R.~A., {Davies}, J.~A., {Rouillard}, A.~P., {et~al.} 2009,
  \solphys, 256, 219

\bibitem[{{Hasan}(2012)}]{hasan2012}
{Hasan}, S.~S. 2012, in ASP Conf. Ser., Vol. 463, Second ATST-EAST Meeting:
  Magnetic Fields from the Photosphere to the Corona., ed. T.~R. {Rimmele},
  A.~{Tritschler}, F.~{W{\"o}ger}, M.~{Collados Vera}, H.~{Socas-Navarro},
  R.~{Schlichenmaier}, M.~{Carlsson}, T.~{Berger}, A.~{Cadavid}, P.~R.
  {Gilbert}, P.~R. {Goode}, \& M.~{Kn{\"o}lker}, 395

\bibitem[{{Hirzberger} {et~al.}(2010){Hirzberger}, {Feller}, {Riethm{\"u}ller},
  {Sch{\"u}ssler}, {Borrero}, {Afram}, {Unruh}, {Berdyugina}, {Gandorfer},
  {Solanki}, {Barthol}, {Bonet}, {Mart{\'{\i}}nez Pillet}, {Berkefeld},
  {Kn{\"o}lker}, {Schmidt}, \& {Title}}]{hirzbergeretal2010}
{Hirzberger}, J., {Feller}, A., {Riethm{\"u}ller}, T.~L., {et~al.} 2010, \apjl,
  723, L154

\bibitem[{{Ishikawa} {et~al.}(2014{\natexlab{a}}){Ishikawa}, {Asensio Ramos},
  {Belluzzi}, {Manso Sainz}, {{\v S}t{\v e}p{\'a}n}, {Trujillo Bueno}, {Goto},
  \& {Tsuneta}}]{ishikawaetal2014}
{Ishikawa}, R., {Asensio Ramos}, A., {Belluzzi}, L., {et~al.}
  2014{\natexlab{a}}, \apj, 787, 159

\bibitem[{{Ishikawa} {et~al.}(2014{\natexlab{b}}){Ishikawa}, {Narukage},
  {Kubo}, {Ishikawa}, {Kano}, \& {Tsuneta}}]{ishikawaetal2014a}
{Ishikawa}, R., {Narukage}, N., {Kubo}, M., {et~al.} 2014{\natexlab{b}},
  \solphys, 289, 4727

\bibitem[{{Jaeggli} {et~al.}(2010){Jaeggli}, {Lin}, {Mickey}, {Kuhn}, {Hegwer},
  {Rimmele}, \& {Penn}}]{jaegglietal2010}
{Jaeggli}, S.~A., {Lin}, H., {Mickey}, D.~L., {et~al.} 2010, \memsai, 81, 763

\bibitem[{{Ji} {et~al.}(2012){Ji}, {Cao}, \& {Goode}}]{jietal2012}
{Ji}, H., {Cao}, W., \& {Goode}, P.~R. 2012, \apjl, 750, L25

\bibitem[{{Kano} {et~al.}(2012){Kano}, {Bando}, {Narukage}, {Ishikawa},
  {Tsuneta}, {Katsukawa}, {Kubo}, {Ishikawa}, {Hara}, {Shimizu}, {Suematsu},
  {Ichimoto}, {Sakao}, {Goto}, {Kato}, {Imada}, {Kobayashi}, {Holloway},
  {Winebarger}, {Cirtain}, {De Pontieu}, {Casini}, {Trujillo Bueno}, {{\v
  S}tep{\'a}n}, {Manso Sainz}, {Belluzzi}, {Asensio Ramos}, {Auch{\`e}re}, \&
  {Carlsson}}]{kanoetal2012}
{Kano}, R., {Bando}, T., {Narukage}, N., {et~al.} 2012, in SPIE Conf. Ser.,
  Vol. 8443, 4

\bibitem[{{Kiselman} {et~al.}(2011){Kiselman}, {Pereira}, {Gustafsson},
  {Asplund}, {Mel{\'e}ndez}, \& {Langhans}}]{kiselmanetal2011}
{Kiselman}, D., {Pereira}, T.~M.~D., {Gustafsson}, B., {et~al.} 2011, \aap,
  535, A14

\bibitem[{{Kleint} {et~al.}(2015){Kleint}, {Battaglia}, {Reardon}, {Sainz
  Dalda}, {Young}, \& {Krucker}}]{kleintetal2015}
{Kleint}, L., {Battaglia}, M., {Reardon}, K., {et~al.} 2015, \apj, 806, 9

\bibitem[{{Kleint} {et~al.}(2011{\natexlab{a}}){Kleint}, {Feller}, \&
  {Gisler}}]{Kleintetal2011}
{Kleint}, L., {Feller}, A., \& {Gisler}, D. 2011{\natexlab{a}}, \aap, 529, A78

\bibitem[{{Kleint} {et~al.}(2011{\natexlab{b}}){Kleint}, {Shapiro},
  {Berdyugina}, \& {Bianda}}]{kleintetal2011b}
{Kleint}, L., {Shapiro}, A.~I., {Berdyugina}, S.~V., \& {Bianda}, M.
  2011{\natexlab{b}}, \aap, 536, A47

\bibitem[{{Kubo} {et~al.}(2014){Kubo}, {Kano}, {Kobayashi}, {Bando},
  {Narukage}, {Ishikawa}, {Tsuneta}, {Katsukawa}, {Ishikawa}, {Suematsu},
  {Hara}, {Shimizu}, {Sakao}, {Ichimoto}, {Goto}, {Holloway}, {Winebarger},
  {Cirtain}, {De Pontieu}, {Casini}, {Auch{\`e}re}, {Trujillo Bueno}, {Manso
  Sainz}, {Belluzzi}, {Asensio Ramos}, {{\v S}t{\v e}p{\'a}n}, \&
  {Carlsson}}]{kuboetal2014}
{Kubo}, M., {Kano}, R., {Kobayashi}, K., {et~al.} 2014, in ASP Conference
  Series, Vol. 489, Solar Polarization 7, ed. K.~N. {Nagendra}, J.~O.
  {Stenflo}, Q.~{Qu}, \& M.~{Samooprna}, 307

\bibitem[{{Lin} {et~al.}(2004){Lin}, {Kuhn}, \& {Coulter}}]{linetal2004}
{Lin}, H., {Kuhn}, J.~R., \& {Coulter}, R. 2004, \apjl, 613, L177

\bibitem[{{Lites} {et~al.}(1993){Lites}, {Elmore}, {Seagraves}, \&
  {Skumanich}}]{litesetal1993}
{Lites}, B.~W., {Elmore}, D.~F., {Seagraves}, P., \& {Skumanich}, A.~P. 1993,
  \apj, 418, 928

\bibitem[{{Lites} {et~al.}(2008){Lites}, {Kubo}, {Socas-Navarro}, {Berger},
  {Frank}, {Shine}, {Tarbell}, {Title}, {Ichimoto}, {Katsukawa}, {Tsuneta},
  {Suematsu}, {Shimizu}, \& {Nagata}}]{litesetal2008}
{Lites}, B.~W., {Kubo}, M., {Socas-Navarro}, H., {et~al.} 2008, \apj, 672, 1237

\bibitem[{{Lites} {et~al.}(1996){Lites}, {Leka}, {Skumanich}, {Martinez
  Pillet}, \& {Shimizu}}]{litesetal1996}
{Lites}, B.~W., {Leka}, K.~D., {Skumanich}, A., {Martinez Pillet}, V., \&
  {Shimizu}, T. 1996, \apj, 460, 1019

\bibitem[{{Liu} {et~al.}(2012){Liu}, {Deng}, {Jin}, \& {Ji}}]{liuetal2012}
{Liu}, Z., {Deng}, Y., {Jin}, Z., \& {Ji}, H. 2012, in SPIE Conf. Ser., Vol.
  8444, 5

\bibitem[{{Liu} {et~al.}(2014{\natexlab{a}}){Liu}, {Jin}, {Yuan}, {Lin},
  {Deng}, {Ji}, \& {Yan}}]{liuetal2014cgst}
{Liu}, Z., {Jin}, Z., {Yuan}, S., {et~al.} 2014{\natexlab{a}}, in SPIE Conf.
  Ser., Vol. 9145, 26

\bibitem[{{Liu} {et~al.}(2014{\natexlab{b}}){Liu}, {Xu}, {Gu}, {Wang}, {You},
  {Shen}, {Lu}, {Jin}, {Chen}, {Lou}, {Li}, {Liu}, {Xu}, {Rao}, {Hu}, {Li},
  {Fu}, {Wang}, {Bao}, {Wu}, \& {Zhang}}]{liuetal2014}
{Liu}, Z., {Xu}, J., {Gu}, B.-Z., {et~al.} 2014{\natexlab{b}}, Research in
  Astronomy and Astrophysics, 14, 705

\bibitem[{{L{\'o}pez Ariste} {et~al.}(2006){L{\'o}pez Ariste}, {Aulanier},
  {Schmieder}, \& {Sainz Dalda}}]{arturoetal2006}
{L{\'o}pez Ariste}, A., {Aulanier}, G., {Schmieder}, B., \& {Sainz Dalda}, A.
  2006, \aap, 456, 725

\bibitem[{{L{\'o}pez Ariste} {et~al.}(2011){L{\'o}pez Ariste}, {Le Men}, \&
  {Gelly}}]{arturo2011tunis}
{L{\'o}pez Ariste}, A., {Le Men}, C., \& {Gelly}, B. 2011, Contributions of the
  Astronomical Observatory Skalnate Pleso, 41, 99

\bibitem[{{L{\'o}pez Ariste} {et~al.}(2010){L{\'o}pez Ariste}, {Le Men},
  {Gelly}, \& {Asensio Ramos}}]{arturo2010tunis}
{L{\'o}pez Ariste}, A., {Le Men}, C., {Gelly}, B., \& {Asensio Ramos}, A. 2010,
  Astronomische Nachrichten, 331, 658

\bibitem[{{L{\'o}pez Ariste} {et~al.}(2012){L{\'o}pez Ariste}, {Leblanc},
  {Casini}, {Manso Sainz}, {Gelly}, \& {Le Men}}]{lopezariste2012mercury}
{L{\'o}pez Ariste}, A., {Leblanc}, F., {Casini}, R., {et~al.} 2012, Icarus,
  220, 1104

\bibitem[{{L{\'o}pez Ariste} {et~al.}(2000){L{\'o}pez Ariste}, {Rayrole}, \&
  {Semel}}]{arturo2000mtr}
{L{\'o}pez Ariste}, A., {Rayrole}, J., \& {Semel}, M. 2000, A\&A Suppl., 142,
  137

\bibitem[{{L{\"o}ptien} {et~al.}(2014){L{\"o}ptien}, {Birch}, {Gizon}, {Schou},
  {Appourchaux}, {Blanco Rodr{\'{\i}}guez}, {Cally}, {Dominguez-Tagle},
  {Gandorfer}, {Hill}, {Hirzberger}, {Scherrer}, \&
  {Solanki}}]{loeptienetal2014}
{L{\"o}ptien}, B., {Birch}, A.~C., {Gizon}, L., {et~al.} 2014, \ssr

\bibitem[{{Mart{\'{\i}}nez Pillet}(2013)}]{valentin2013}
{Mart{\'{\i}}nez Pillet}, V. 2013, \ssr, 178, 141

\bibitem[{{Mart{\'{\i}}nez Pillet} {et~al.}(2011){Mart{\'{\i}}nez Pillet}, {Del
  Toro Iniesta}, {{\'A}lvarez-Herrero}, {Domingo}, {Bonet}, {Gonz{\'a}lez
  Fern{\'a}ndez}, {L{\'o}pez Jim{\'e}nez}, {Pastor}, {Gasent Blesa}, {Mellado},
  {Piqueras}, {Aparicio}, {Balaguer}, {Ballesteros}, {Belenguer}, {Bellot
  Rubio}, {Berkefeld}, {Collados}, {Deutsch}, {Feller}, {Girela}, {Grauf},
  {Heredero}, {Herranz}, {Jer{\'o}nimo}, {Laguna}, {Meller}, {Men{\'e}ndez},
  {Morales}, {Orozco Su{\'a}rez}, {Ramos}, {Reina}, {Ramos},
  {Rodr{\'{\i}}guez}, {S{\'a}nchez}, {Uribe-Patarroyo}, {Barthol}, {Gandorfer},
  {Knoelker}, {Schmidt}, {Solanki}, \& {Vargas
  Dom{\'{\i}}nguez}}]{valentinetal2011}
{Mart{\'{\i}}nez Pillet}, V., {Del Toro Iniesta}, J.~C., {{\'A}lvarez-Herrero},
  A., {et~al.} 2011, \solphys, 268, 57

\bibitem[{{Mein}(2002)}]{mein2002}
{Mein}, P. 2002, \aap, 381, 271

\bibitem[{{Orozco Su{\'a}rez} \& {Bellot Rubio}(2012)}]{orozcobellot2012}
{Orozco Su{\'a}rez}, D. \& {Bellot Rubio}, L.~R. 2012, \apj, 751, 2

\bibitem[{{Pesnell} {et~al.}(2012){Pesnell}, {Thompson}, \&
  {Chamberlin}}]{sdo2012}
{Pesnell}, W.~D., {Thompson}, B.~J., \& {Chamberlin}, P.~C. 2012, \solphys,
  275, 3

\bibitem[{{Peter} {et~al.}(2012){Peter}, {Abbo}, {Andretta}, {Auch{\`e}re},
  {Bemporad}, {Berrilli}, {Bommier}, {Braukhane}, {Casini}, {Curdt}, {Davila},
  {Dittus}, {Fineschi}, {Fludra}, {Gandorfer}, {Griffin}, {Inhester}, {Lagg},
  {Degl'Innocenti}, {Maiwald}, {Sainz}, {Pillet}, {Matthews}, {Moses},
  {Parenti}, {Pietarila}, {Quantius}, {Raouafi}, {Raymond}, {Rochus},
  {Romberg}, {Schlotterer}, {Sch{\"u}hle}, {Solanki}, {Spadaro}, {Teriaca},
  {Tomczyk}, {Bueno}, \& {Vial}}]{peteretal2012}
{Peter}, H., {Abbo}, L., {Andretta}, V., {et~al.} 2012, Experimental Astron.,
  33, 271

\bibitem[{{Povel}(1995)}]{Povel1995}
{Povel}, H.-P. 1995, Optical Engineering, 34, 1870

\bibitem[{{Puschmann} {et~al.}(2013){Puschmann}, {Denker}, {Balthasar},
  {Louis}, {Popow}, {Woche}, {Beck}, {Seelemann}, \&
  {Volkmer}}]{gregorgfpi2013}
{Puschmann}, K.~G., {Denker}, C., {Balthasar}, H., {et~al.} 2013, Optical
  Engineering, 52, 081606

\bibitem[{{Raja Bayanna} {et~al.}(2014){Raja Bayanna}, {Mathew},
  {Venkatakrishnan}, \& {Srivastava}}]{bayannaetal2014}
{Raja Bayanna}, A., {Mathew}, S.~K., {Venkatakrishnan}, P., \& {Srivastava}, N.
  2014, \solphys, 289, 4007

\bibitem[{{Ramelli} {et~al.}(2014){Ramelli}, {Gisler}, {Bianda}, {Bello
  Gonz{\'a}lez}, {Berdyugina}, \& {Soltau}}]{gregorzimpol2014}
{Ramelli}, R., {Gisler}, D., {Bianda}, M., {et~al.} 2014, in SPIE Conf. Ser.,
  Vol. 9147, 3

\bibitem[{{Rempel} \& {Schlichenmaier}(2011)}]{rempelschlichenmaier2011}
{Rempel}, M. \& {Schlichenmaier}, R. 2011, Living Reviews in Solar Physics, 8,
  3

\bibitem[{{Riethm{\"u}ller} {et~al.}(2010){Riethm{\"u}ller}, {Solanki},
  {Mart{\'{\i}}nez Pillet}, {Hirzberger}, {Feller}, {Bonet}, {Bello
  Gonz{\'a}lez}, {Franz}, {Sch{\"u}ssler}, {Barthol}, {Berkefeld}, {del Toro
  Iniesta}, {Domingo}, {Gandorfer}, {Kn{\"o}lker}, \&
  {Schmidt}}]{riethmuelleretal2010}
{Riethm{\"u}ller}, T.~L., {Solanki}, S.~K., {Mart{\'{\i}}nez Pillet}, V.,
  {et~al.} 2010, \apjl, 723, L169

\bibitem[{{Rimmele} {et~al.}(2004){Rimmele}, {Richards}, {Hegwer}, {Fletcher},
  {Gregory}, {Moretto}, {Didkovsky}, {Denker}, {Dolgushin}, {Goode},
  {Langlois}, {Marino}, \& {Marquette}}]{rimmeleetal2004}
{Rimmele}, T.~R., {Richards}, K., {Hegwer}, S., {et~al.} 2004, in SPIE Conf.
  Ser., Vol. 5171, Telescopes and Instrumentation for Solar Astrophysics, ed.
  S.~{Fineschi} \& M.~A. {Gummin}, 179--186

\bibitem[{{Romano} {et~al.}(2014){Romano}, {Guglielmino}, {Cristaldi},
  {Ermolli}, {Falco}, \& {Zuccarello}}]{romanoetal2014}
{Romano}, P., {Guglielmino}, S.~L., {Cristaldi}, A., {et~al.} 2014, \apj, 784,
  10

\bibitem[{{Sankarasubramanian} {et~al.}(2004){Sankarasubramanian}, {Gullixson},
  {Hegwer}, {Rimmele}, {Gregory}, {Spence}, {Fletcher}, {Richards}, {Rousset},
  {Lites}, {Elmore}, {Streander}, \& {Sigwarth}}]{sankarasubramanian2004}
{Sankarasubramanian}, K., {Gullixson}, C., {Hegwer}, S., {et~al.} 2004, in SPIE
  Conf. Ser., Vol. 5171, Telescopes and Instrumentation for Solar Astrophysics,
  ed. S.~{Fineschi} \& M.~A. {Gummin}, 207--218

\bibitem[{{Schad} {et~al.}(2013){Schad}, {Penn}, \& {Lin}}]{schadetal2013}
{Schad}, T.~A., {Penn}, M.~J., \& {Lin}, H. 2013, \apj, 768, 111

\bibitem[{{Scharmer} {et~al.}(2003){Scharmer}, {Bjelksjo}, {Korhonen},
  {Lindberg}, \& {Petterson}}]{scharmeretal2003}
{Scharmer}, G.~B., {Bjelksjo}, K., {Korhonen}, T.~K., {Lindberg}, B., \&
  {Petterson}, B. 2003, in SPIE Conf. Ser., Vol. 4853, Innovative Telescopes
  and Instrumentation for Solar Astrophysics, ed. S.~L. {Keil} \& S.~V.
  {Avakyan}, 341--350

\bibitem[{{Scharmer} {et~al.}(2013){Scharmer}, {de la Cruz Rodriguez},
  {S{\"u}tterlin}, \& {Henriques}}]{scharmeretal2013}
{Scharmer}, G.~B., {de la Cruz Rodriguez}, J., {S{\"u}tterlin}, P., \&
  {Henriques}, V.~M.~J. 2013, \aap, 553, A63

\bibitem[{{Scharmer} {et~al.}(2008){Scharmer}, {Narayan}, {Hillberg}, {de la
  Cruz Rodr{\'{\i}}guez}, {L{\"o}fdahl}, {Kiselman}, {S{\"u}tterlin}, {van
  Noort}, \& {Lagg}}]{scharmeretal2008}
{Scharmer}, G.~B., {Narayan}, G., {Hillberg}, T., {et~al.} 2008, \apjl, 689,
  L69

\bibitem[{{Schmidt} {et~al.}(2012){Schmidt}, {von der L{\"u}he}, {Volkmer},
  {Denker}, {Solanki}, {Balthasar}, {Bello Gonzalez}, {Berkefeld}, {Collados},
  {Fischer}, {Halbgewachs}, {Heidecke}, {Hofmann}, {Kneer}, {Lagg}, {Nicklas},
  {Popow}, {Puschmann}, {Schmidt}, {Sigwarth}, {Sobotka}, {Soltau}, {Staude},
  {Strassmeier}, \& {Waldmann }}]{gregor2012an}
{Schmidt}, W., {von der L{\"u}he}, O., {Volkmer}, R., {et~al.} 2012,
  Astronomische Nachrichten, 333, 796

\bibitem[{{Schmieder} {et~al.}(2014){Schmieder}, {Tian}, {Kucera}, {L{\'o}pez
  Ariste}, {Mein}, {Mein}, {Dalmasse}, \& {Golub}}]{schmiederetal2014}
{Schmieder}, B., {Tian}, H., {Kucera}, T., {et~al.} 2014, \aap, 569, A85

\bibitem[{{Shapiro} {et~al.}(2011){Shapiro}, {Fluri}, {Berdyugina}, {Bianda},
  \& {Ramelli}}]{shapiroetal2011}
{Shapiro}, A.~I., {Fluri}, D.~M., {Berdyugina}, S.~V., {Bianda}, M., \&
  {Ramelli}, R. 2011, \aap, 529, A139

\bibitem[{{Socas-Navarro}(2005)}]{socasnavarro2005}
{Socas-Navarro}, H. 2005, \apjl, 631, L167

\bibitem[{{Socas-Navarro} {et~al.}(2008){Socas-Navarro}, {Borrero}, {Asensio
  Ramos}, {Collados}, {Dom{\'{\i}}nguez Cerde{\~n}a}, {Khomenko},
  {Mart{\'{\i}}nez Gonz{\'a}lez}, {Mart{\'{\i}}nez Pillet}, {Ruiz Cobo}, \&
  {S{\'a}nchez Almeida}}]{socasnavarroetal2008}
{Socas-Navarro}, H., {Borrero}, J.~M., {Asensio Ramos}, A., {et~al.} 2008,
  \apj, 674, 596

\bibitem[{{Socas-Navarro} \& {Elmore}(2005)}]{socasnavarroelmore2005}
{Socas-Navarro}, H. \& {Elmore}, D. 2005, \apjl, 619, L195

\bibitem[{{Socas-Navarro} {et~al.}(2006){Socas-Navarro}, {Elmore}, {Pietarila},
  {Darnell}, {Lites}, {Tomczyk}, \& {Hegwer}}]{socasnavarroetal2006}
{Socas-Navarro}, H., {Elmore}, D., {Pietarila}, A., {et~al.} 2006, \solphys,
  235, 55

\bibitem[{{Solanki} \& {the EPIC consortium}(2015)}]{solankietal2015prop}
{Solanki}, S. \& {the EPIC consortium}. 2015, Proposal for M4 Mission of ESA

\bibitem[{{Solanki} {et~al.}(2011){Solanki}, {Barthol}, {Danilovic}, {Feller},
  {Gandorfer}, {Hirzberger}, {Lagg}, {Riethm{\"u}ller}, {Sch{\"u}ssler},
  {Wiegelmann}, {Bonet}, {Pillet}, {Khomenko}, {del Toro Iniesta}, {Domingo},
  {Palacios}, {Kn{\"o}lker}, {Gonz{\'a}lez}, {Borrero}, {Berkefeld}, {Franz},
  {Roth}, {Schmidt}, {Steiner}, \& {Title}}]{solankietal2011}
{Solanki}, S.~K., {Barthol}, P., {Danilovic}, S., {et~al.} 2011, in IAU
  Symposium, ed. D.~{Prasad Choudhary} \& K.~G. {Strassmeier}, Vol. 273,
  226--232

\bibitem[{{Solanki} {et~al.}(2010){Solanki}, {Barthol}, {Danilovic}, {Feller},
  {Gandorfer}, {Hirzberger}, {Riethm{\"u}ller}, {Sch{\"u}ssler}, {Bonet},
  {Mart{\'{\i}}nez Pillet}, {del Toro Iniesta}, {Domingo}, {Palacios},
  {Kn{\"o}lker}, {Bello Gonz{\'a}lez}, {Berkefeld}, {Franz}, {Schmidt}, \&
  {Title}}]{solankietal2010}
{Solanki}, S.~K., {Barthol}, P., {Danilovic}, S., {et~al.} 2010, \apjl, 723,
  L127

\bibitem[{{Solanki} {et~al.}(2015){Solanki}, {del Toro Iniesta}, {Woch},
  {Gandorfer}, {Hirzberger}, {Schmidt}, {Appourchaux}, \&
  {Alvarez-Herrero}}]{solankietal2015a}
{Solanki}, S.~K., {del Toro Iniesta}, J.~C., {Woch}, J., {et~al.} 2015, in IAU
  Symposium, Vol. 305, IAU Symposium, 108--113

\bibitem[{{Steiner} \& {Rezaei}(2012)}]{steinerrezaei2012}
{Steiner}, O. \& {Rezaei}, R. 2012, in ASP Conf. Ser., Vol. 456, Fifth Hinode
  Science Meeting, ed. L.~{Golub}, I.~{De Moortel}, \& T.~{Shimizu}, 3

\bibitem[{{Stenflo}(2010)}]{stenflo2010}
{Stenflo}, J.~O. 2010, \aap, 517, A37

\bibitem[{{Stenflo} {et~al.}(1976){Stenflo}, {Biverot}, \&
  {Stenmark}}]{stenfloetal1976}
{Stenflo}, J.~O., {Biverot}, H., \& {Stenmark}, L. 1976, Appl. Optics, 15, 1188

\bibitem[{{Stenflo} {et~al.}(1980){Stenflo}, {Dravins}, {Wihlborg}, {Bruns},
  {Prokofev}, {Zhitnik}, {Biverot}, \& {Stenmark}}]{stenfloetal1980}
{Stenflo}, J.~O., {Dravins}, D., {Wihlborg}, N., {et~al.} 1980, \solphys, 66,
  13

\bibitem[{{Stenflo} \& {Keller}(1997)}]{stenflokeller1997}
{Stenflo}, J.~O. \& {Keller}, C.~U. 1997, \aap, 321, 927

\bibitem[{{Stenflo} \& {Stenholm}(1976)}]{stenflostenholm1976}
{Stenflo}, J.~O. \& {Stenholm}, L. 1976, \aap, 46, 69

\bibitem[{{Suematsu} {et~al.}(2014{\natexlab{a}}){Suematsu}, {Katsukawa},
  {Hara}, {Kano}, {Shimizu}, \& {Ichimoto}}]{suematsuetal2014solarc}
{Suematsu}, Y., {Katsukawa}, Y., {Hara}, H., {et~al.} 2014{\natexlab{a}}, in
  SPIE Conf. Ser., Vol. 9143, 1

\bibitem[{{Suematsu} {et~al.}(2014{\natexlab{b}}){Suematsu}, {Sukegawa},
  {Okura}, {Nakayasu}, {Enokida}, {Koyama}, {Saito}, {Ozaki}, \&
  {Tsuneta}}]{suematsuetal2014slicer}
{Suematsu}, Y., {Sukegawa}, T., {Okura}, Y., {et~al.} 2014{\natexlab{b}}, in
  SPIE Conf. Ser., Vol. 9151, 1

\bibitem[{{Suematsu} {et~al.}(2008){Suematsu}, {Tsuneta}, {Ichimoto},
  {Shimizu}, {Otsubo}, {Katsukawa}, {Nakagiri}, {Noguchi}, {Tamura}, {Kato},
  {Hara}, {Kubo}, {Mikami}, {Saito}, {Matsushita}, {Kawaguchi}, {Nakaoji},
  {Nagae}, {Shimada}, {Takeyama}, \& {Yamamuro}}]{suematsuetal2008}
{Suematsu}, Y., {Tsuneta}, S., {Ichimoto}, K., {et~al.} 2008, \solphys, 249,
  197

\bibitem[{{Teriaca} {et~al.}(2012){Teriaca}, {Andretta}, {Auch{\`e}re},
  {Brown}, {Buchlin}, {Cauzzi}, {Culhane}, {Curdt}, {Davila}, {Del Zanna},
  {Doschek}, {Fineschi}, {Fludra}, {Gallagher}, {Green}, {Harra}, {Imada},
  {Innes}, {Kliem}, {Korendyke}, {Mariska}, {Mart{\'{\i}}nez-Pillet},
  {Parenti}, {Patsourakos}, {Peter}, {Poletto}, {Rutten}, {Sch{\"u}hle},
  {Siemer}, {Shimizu}, {Socas-Navarro}, {Solanki}, {Spadaro}, {Trujillo-Bueno},
  {Tsuneta}, {Dominguez}, {Vial}, {Walsh}, {Warren}, {Wiegelmann}, {Winter}, \&
  {Young}}]{teriacaetal2012}
{Teriaca}, L., {Andretta}, V., {Auch{\`e}re}, F., {et~al.} 2012, Experimental
  Astronomy, 34, 273

\bibitem[{{Thomas} {et~al.}(1982){Thomas}, {Cram}, \& {Nye}}]{thomasetal1982}
{Thomas}, J.~H., {Cram}, L.~E., \& {Nye}, A.~H. 1982, \nat, 297, 485

\bibitem[{{Trujillo Bueno}(2003)}]{trujillobueno2003}
{Trujillo Bueno}, J. 2003, in Astronomical Society of the Pacific Conference
  Series, Vol. 307, Solar Polarization, ed. J.~{Trujillo-Bueno} \& J.~{Sanchez
  Almeida}, 407

\bibitem[{{Trujillo Bueno} {et~al.}(2006){Trujillo Bueno}, {Asensio Ramos}, \&
  {Shchukina}}]{trujillobuenoetal2006}
{Trujillo Bueno}, J., {Asensio Ramos}, A., \& {Shchukina}, N. 2006, in ASP
  Conf. Ser., ed. R.~{Casini} \& B.~W. {Lites}, Vol. 358, 269

\bibitem[{{Trujillo Bueno} {et~al.}(2011){Trujillo Bueno}, {{\v S}t{\v
  e}p{\'a}n}, \& {Casini}}]{trujillobuenoetal2011}
{Trujillo Bueno}, J., {{\v S}t{\v e}p{\'a}n}, J., \& {Casini}, R. 2011, \apjl,
  738, L11

\bibitem[{{{\v S}t{\v e}p{\'a}n}(2015)}]{stepan2015}
{{\v S}t{\v e}p{\'a}n}, J. 2015, in IAU Symposium, Vol. 305, IAU Symposium,
  360--367

\bibitem[{{{\v S}t{\v e}p{\'a}n} {et~al.}(2015){{\v S}t{\v e}p{\'a}n},
  {Trujillo Bueno}, {Leenaarts}, \& {Carlsson}}]{stepanetal2015}
{{\v S}t{\v e}p{\'a}n}, J., {Trujillo Bueno}, J., {Leenaarts}, J., \&
  {Carlsson}, M. 2015, \apj, 803, 65

\bibitem[{{van Noort} {et~al.}(2005){van Noort}, {Rouppe van der Voort}, \&
  {L{\"o}fdahl}}]{vannoortetal2005}
{van Noort}, M., {Rouppe van der Voort}, L., \& {L{\"o}fdahl}, M.~G. 2005,
  \solphys, 228, 191

\bibitem[{{Vargas Dom{\'{\i}}nguez} {et~al.}(2014){Vargas Dom{\'{\i}}nguez},
  {Kosovichev}, \& {Yurchyshyn}}]{vargasetal2014}
{Vargas Dom{\'{\i}}nguez}, S., {Kosovichev}, A., \& {Yurchyshyn}, V. 2014,
  \apj, 794, 140

\bibitem[{{Varsik} {et~al.}(2014){Varsik}, {Plymate}, {Goode}, {Kosovichev},
  {Cao}, {Coulter}, {Ahn}, {Gorceix}, \& {Shumko}}]{varsikbbso2014}
{Varsik}, J., {Plymate}, C., {Goode}, P., {et~al.} 2014, in SPIE Conf. Ser.,
  Vol. 9147, 5

\bibitem[{Wang {et~al.}(2015)Wang, Cao, Liu, Xu, Liu, Zeng, Chae, \&
  Ji}]{wangetal2015}
Wang, H., Cao, W., Liu, C., {et~al.} 2015, Nat Commun, 6, 7008

\bibitem[{{Wang} {et~al.}(2014){Wang}, {Liu}, {Deng}, {Zeng}, {Xu}, {Jing}, \&
  {Cao}}]{wangetal2014}
{Wang}, H., {Liu}, C., {Deng}, N., {et~al.} 2014, \apjl, 781, L23

\bibitem[{{Watanabe}(2014)}]{watanabe2014}
{Watanabe}, T. 2014, in SPIE Conf. Ser., Vol. 9143, 1

\bibitem[{{Yang} {et~al.}(2014){Yang}, {Zhang}, {Liu}, \&
  {Xiang}}]{yangetal2014}
{Yang}, S., {Zhang}, J., {Liu}, Z., \& {Xiang}, Y. 2014, \apjl, 784, L36

\bibitem[{{Yurchyshyn} {et~al.}(2015){Yurchyshyn}, {Abramenko}, \&
  {Kilcik}}]{yurchyshynetal2015}
{Yurchyshyn}, V., {Abramenko}, V., \& {Kilcik}, A. 2015, \apj, 798, 136

\bibitem[{{Zeng} {et~al.}(2013){Zeng}, {Cao}, \& {Ji}}]{zengetal2013}
{Zeng}, Z., {Cao}, W., \& {Ji}, H. 2013, \apjl, 769, L33

\bibitem[{{Zeng} {et~al.}(2014){Zeng}, {Qiu}, {Cao}, \& {Judge}}]{zengetal2014}
{Zeng}, Z., {Qiu}, J., {Cao}, W., \& {Judge}, P.~G. 2014, \apj, 793, 87

\bibitem[{{Zirker}(1998)}]{zirker1998}
{Zirker}, J.~B. 1998, \solphys, 182, 1

\end{thebibliography}

\end{document}